\documentclass[a4paper,11pt]{article}
\pdfoutput=1
\usepackage{jcappub}

\usepackage{graphicx}
\usepackage{color}
\usepackage[dvipsnames]{xcolor}
\usepackage{xspace}
\usepackage{multirow}
\usepackage{cleveref}
\usepackage[normalem]{ulem}
\usepackage{makecell}

\newcommand{\tquote}[1]{``#1''}

\def\lb{LiteBIRD}
\def\sfour{{CMB-S4}}
\newcommand{\lnAs}{\ln(10^{10} A_{\rm s})}

\author[1]{Hao Fu,}
\author[2]{Matteo Lucca,}
\author[3]{Silvia Galli,}
\author[4]{Elia S. Battistelli,}
\author[2]{Deanna C. Hooper,}
\author[5]{Julien Lesgourgues,}
\author[5]{Nils Sch\"{o}neberg}

\affiliation[1]{Department of Physics and Astronomy, University of Southampton, \\ Highfield Campus, SO17 1BJ, UK}
\affiliation[2]{Service de Physique Th\'{e}orique, Universit\'{e} Libre de Bruxelles, \\ C.P. 225, B-1050 Brussels, Belgium}
\affiliation[3]{Sorbonne Universit\'e, CNRS, UMR 7095, Institut d'Astrophysique de Paris, \\ 98 bis bd Arago, 75014 Paris, France}
\affiliation[4]{Physics department, Sapienza University of Rome, \\ Piazzale Aldo Moro 5, 00185, Rome, Italy}
\affiliation[5]{Institute for Theoretical Particle Physics and Cosmology (TTK), RWTH Aachen University, \\ D-52056 Aachen, Germany}

\emailAdd{h.fu@soton.ac.uk}
\emailAdd{mlucca@ulb.ac.be} 
\emailAdd{gallis@iap.fr} 

\title{Unlocking the synergy between CMB spectral distortions and anisotropies}

\abstract{
    Measurements of the cosmic microwave background (CMB) spectral distortions (SDs) will open a new window on the very early universe, providing new information complementary to that gathered from CMB temperature and polarization anisotropies. In this paper, we study their synergy as a function of the characteristics of the considered experiments. In particular, we examine a wide range of sensitivities for possible SD measurements, spanning from FIRAS up to noise levels 1000 times better than PIXIE, and study their constraining power when combined with current or future CMB anisotropy experiments such as Planck or \lb\ plus \sfour. We consider a number of different cosmological models such as the $\Lambda$CDM, as well as its extensions with the running of the scalar spectral index, the decay or the annihilation of dark matter (DM) particles. While upcoming CMB anisotropy experiments will be able to  decrease the uncertainties on inflationary parameters such as $A_s$ and $n_s$ by about a factor 2 in the $\Lambda$CDM case, we find that an SD experiment 100 times more sensitive than PIXIE (comparable to the proposed Super-PIXIE satellite) could potentially further contribute to constrain these parameters. This is even more significant in the  case of the running of the scalar spectral index. Furthermore, as expected, constraints on DM particles decaying at redshifts probed by SDs will improve by orders of magnitude even with an experiment 10 times worse than PIXIE as compared to CMB anisotropies or Big Bang Nucleosynthesis bounds. On the contrary, DM annihilation constraints will not significantly improve over CMB anisotropy measurements. Finally, we forecast the constraints obtainable with sensitivities achievable either from the ground or from a balloon.
}

\begin{document}

\hfill{\small ULB-TH/20-06}\\
\vspace{-0.5 cm}
\hfill{\small TTK-20-18}

\maketitle

\section{Introduction}
The cosmic microwave background (CMB) is a remarkably rich source of information about early universe physics. In particular during the last decades significant effort has been devoted to the investigation of the CMB temperature and polarization anisotropies, including also the latest data release of the ESA Planck mission \cite{planck_2018_cosmo_params} (from now on referred to as Planck), which allowed to build a very accurate all-sky map and the power spectrum of the CMB.

Another very promising and not yet fully explored complementary cosmological probe is given by CMB spectral distortions (SDs) (see e.g. \cite{Chluba:2019kpb, lucca_2019} for recent reviews). According to their spectral shape, these can be classified as a combination of a temperature-shift, a chemical potential $\mu$-type, a Compton $y$-type, and a residual distortion (see e.g. \cite{chluba_jeong_2014} for more details). Indeed, studying distortions of the CMB black body (BB) spectrum allows us to constrain several energy release mechanisms occurring at redshifts $z \lesssim 2 \times 10^6$. A variety of such astrophysical and cosmological processes are expected within the Standard Model of cosmology, while others, if present at all, could be caused by any exotic mechanism injecting energy in the CMB photon bath \cite{sunyaev_zeldovich_1969, sunyaev_zeldovich_1970, illarionov_1975, danese_1990, hu_silk_1993,chluba_sunyaev_2012}.

However, although major progress has been made in the theoretical and numerical prediction of SDs in the last years, exactly three decades have passed since the development of experiments that measured the CMB frequency spectrum, such as the COBRA \cite{gush_1990} and FIRAS \cite{mather_1994, fixsen_1996} missions, which measured the CMB spectrum to be that of a pure BB up to uncertainties in the brightness intensity of $\Delta I_\nu  / I_\nu \lesssim 10^{-5}$. Today, thanks to new technologies, much more accurate SD measurements are feasible and could easily achieve sensitivities three or four orders of magnitude better than those of FIRAS, as for instance in the case of spectrometers like PRISTINE, PIXIE \cite{kogut_2011}, PRISM \cite{prism_2014}, Super-PIXIE \cite{Kogut:2019vqh} and Voyage 2050 \cite{Chluba:2019nxa}, as well as with imaging telescopes as recently proposed in \cite{Mukherjee:2019pcq}.  

Therefore, precise forecasts of the constraining power of different SD observational setups are needed, and addressing this necessity is precisely the goal of this paper. In particular, as a first systematic effort in this direction, we consider a series of experimental designs assuming an idealised foreground removal to investigate the maximum amount of information that could be extracted from SDs on the early universe for a variety of different cosmological models. In this sense, our results are not meant to realistically estimate the reach of different experimental setups (which are themselves anyway idealized to a large extent in our treatment), but rather to study and compare the dependence of the cosmological constraints on the sensitivity of those experiments. For our forecasts we use the latest versions of the Boltzmann solver \textsc{class} \cite{lesgourgues_class_ii} and the Monte Carlo Markov Chain (MCMC) sampler \textsc{MontePython} \cite{audren_2013, brinckmann_2019} as described in~\cite{lucca_2019}, which now allow us to include SD experiments in combination with CMB anisotropy surveys. 

This paper extends some previous work by \cite{lucca_2019} in mainly two directions. On the one hand, we consider several additional SD experiment designs in order to study the minimum sensitivity required to improve constraints on specific cosmological scenarios with respect to the ones provided by CMB anisotropies. On the other hand, we marginalize over the contribution from late-time sources of SDs such as clusters of galaxies, while \cite{lucca_2019} only assumed SDs produced in the early universe.

Indeed, within this work, we forecast the sensitivity of various possible SD experiments other than just PIXIE and PRISM to the cosmological parameters that describe possible sources of energy release. For different cosmological scenarios and a few assumptions concerning future CMB anisotropy experiments, we show how the interplay between CMB SD and anisotropy data could break degeneracies between many of the cosmological parameters involved, especially in the case of the most futuristic SD configurations. This level of synergy between the two probes was out of reach for the low-sensitivity SD configurations considered in previous works, like e.g. \cite{lucca_2019}.

We also present for the first time some MCMC forecasts for experiments attempting to measure SDs from the ground, and we show the improvement that might be achieved with future measurements from the stratosphere or from space. Basing ourselves on existing concept studies for ground-based experiments such as COSMO \cite{COSMO} and BISOU (see e.g.~\cite{Chluba:2019nxa} for additional discussions), we consider the case of an experiment with two frequency bands dictated by the favorable atmospheric window and sensitivity-limited by the photon background from the atmosphere and the CMB. This represents a first attempt to estimate the potential constraining power of these experiments on cosmological parameters. 

For all of the considered cases we account for the fact that primordial $y$ distortions will be hardly distinguishable from the contribution of late-time $y$ sources such as cosmic reionization, the intracluster medium (ICM) of groups and clusters of galaxies, and the intergalactic medium (IGM) between halos~\cite{hill_2015}.  We effectively model these contributions by marginalizing over the optical depth and mean temperature of the dominant late-time source, the ICM. Moreover, we also marginalize over the uncertainty on the monopole temperature $T_0$. However, we assume a perfect removal of some major potential sources of contamination, such as galactic and extra-galactic foregrounds. Also, when considering ground-based or balloon-borne experiments, we also neglect the contribution of atmospheric brightness fluctuations (while we do include the increase in photon noise due to the atmospheric emission).

This paper is organized as follows. In Section \ref{sec_scenarios} we provide an overview on the energy injection mechanisms considered in this work. In Section \ref{sec_exp_config} we describe the specifics of the considered experiments and in Section \ref{sec_method} we describe the method used to forecast their constraining power on cosmological parameters. In Section \ref{sec_res} we display the results of our investigation. Finally, a summary of the results together with additional discussions is given in Section~\ref{sec_concl}.

\section{Energy release scenarios}\label{sec_scenarios}
In our analysis we consider different energy injection mechanisms that can lead to SDs in the CMB energy spectrum focusing exclusively on the \tquote{primordial} contributions (i.e., those that take place prior to recombination, neglecting late-time effects such as the Sunyeav-Zeldovich effect which are instead marginalized over as explained in Sec. \ref{sec_montepython}). Among these processes, we focus on the standard $\Lambda$CDM ones, such as the dissipation of acoustic waves and adiabatic cooling, as well as the annihilation and decay of relic particles. These cases are particularly interesting as they encompass a variety of non-standard scenarios. For instance, in the case of acoustic wave dissipation it is possible to test many alternative inflationary models via the dependence on the primordial power spectrum (see e.g. \cite{Chluba:2012we, Cabass:2016giw, Schoneberg:2020nyg}). In the case of decaying relic particles (see \text{e.g.~\cite{chen_kamionkowski_2004, Poulin:2016anj,Acharya:2019owx}}), the discussion on the required detector sensitivity can be extended to the evaporation of primordial black holes (see e.g. \cite{Poulin:2016anj, Stocker:2018avm}) due to the similar injection histories (see e.g. Figures 4 and 5 of \cite{lucca_2019}). Other possible sources of SDs include interactions between dark matter (DM) and Standard Model particles (see e.g. \cite{Ali-Haimoud:2015pwa, Slatyer:2018aqg, Ali-Haimoud:2021lka}), but since results for one model do not necessarily generalize, we leave the analysis of these different models for future work.

Although the theory underlying these processes is already widely documented in the literature, in this section we provide a brief overview for the sake of completeness. The adopted notation is based on the work of \cite{lucca_2019}, where the interested reader can find more in-depth discussions. Other related works are e.g. \cite{chluba_sunyaev_2012, chluba_jeong_2014}.

\subsection{The $\Lambda$CDM model and running of the spectral index}\label{sec_th_LCDM}

In the $\Lambda$CDM model, we consider two main sources of SDs: the damping of acoustic waves and the baryon adiabatic cooling.

When the primordial energy density perturbations enter the horizon, pressure gradients form and cause pressure waves. The oscillation of these waves (referred to hereafter as acoustic waves) is affected by dissipation, which causes damping at small scales and creates distortions in the CMB frequency spectrum \cite{sunyaev_zeldovich_1970, hu_1994}. The type of distortions that can be generated depends on the epochs of the damping, while the intensity of the signal depends on the amplitude of the damped wave. It is important to specify that the acoustic wave damping is not an energy injection to the CMB radiation field, but rather a redistribution of the radiation field energy (and hence referred to specifically as $\dot{\mathcal Q}_{\text {non }-\mathrm{inj}}$, where $\dot{\mathcal{Q}}$ is the heating rate).

The distortion generated by the damping of the CMB small-scale fluctuations depends on the amplitude and the shape of the primordial power spectrum at scales $1 \text{ Mpc}^{-1} \leq k \leq 2 \times 10^4 \text{ Mpc}^{-1}$ \cite{chluba_sunyaev_2012, chluba_2013, chluba_2016}. An accurate approximation for the effective heating rate and the curvature power spectrum of scalar perturbations is given by \cite{chluba_khatri_2012, chluba_grin_2013}
\begin{equation}
 \dot{\mathcal Q}_{\text {non }-\mathrm{inj}}=4 A^{2} \rho_{\gamma} \partial_z k_\text{D}^{-2} \int_{k_\text{min}}^\infty \frac{k^4 \mathrm dk} {2\pi^2} {\cal P}_{\cal R}(k) \mathrm e^{-2k^2 / k_\text{D}^2}\,,
 \label{eq_diss_ac_waves}
\end{equation}
where $A$ is a normalization factor, $k_\text{D}$ is the photon damping scale \cite{silk_68,kaiser_1983} and ${\cal P}_{\cal R}(k)$ is the dimensionless primordial power spectrum of curvature fluctuations. Alternative approximations can be found e.g. in \cite{Hu:1995em}.
	
On the other hand, baryon adiabatic cooling is caused by the fact that the temperature of photons cools due to the expansion of the universe as $T_\gamma\propto (1+z)$, while baryons cool faster, $T_e\propto (1+z)^2$. Thus, photons transfer energy to baryons when strictly coupled to them in the early universe to maintain equilibrium. This subtraction of energy from the photon field causes distortions in the CMB spectrum, which partially cancel out those due to heating (see e.g. \cite{chluba_sunyaev_2012}).

Within this work, we will consider the standard $\Lambda$CDM case, as well as a minimal extension of the standard primordial power spectrum including the running of the spectral index, i.e.
\begin{equation}
	{\cal P}_{\cal R} (k) = 2 \pi^2 A_\text{s} k^{-3} \left( \frac{k}{k_0} \right) ^ {n_\text{s} -1 + \frac{1}{2} n_\text{run} \ln (k/k_0)},
	\label{eq_power_spectrum}
\end{equation}
where $A_\text{s}$, $n_\text{s}$ and $n_\text{run}$ are the amplitude, the power index and its running, respectively, while $k_0$ is the pivot scale, which we assume to be $k_0=0.05$~Mpc$^{-1}$. The running is set to zero when considering the $\Lambda$CDM case. Overall, the final set of free parameters involved in the $\Lambda$CDM scenario is
\begin{equation}\label{eq_param_set_LCDM}
    \{ \omega_\text{b}, \omega_\text{cdm}, 100 \theta_s, \lnAs, n_\text{s}, \tau_\text{reio} \},
\end{equation}
while for the $\Lambda$CDM+$n_\text{run}$ case is
\begin{equation}\label{eq_param_set_run}
    \{ \omega_\text{b}, \omega_\text{cdm}, 100 \theta_s, \lnAs, n_\text{s}, \tau_\text{reio} \} + n_\text{run}\,,
\end{equation}
with $\omega_\text{b}$ and  $\omega_\text{cdm}$ as the physical baryon and DM densities respectively, $\theta_s$ as the angular scale of sound horizon at last scattering and $\tau_\text{reio}$ as the reionization optical depth.

Using the Planck $\Lambda$CDM best-fit model \cite{planck_2018_cosmo_params}, one can predict the expected amplitudes of the $y$ and $\mu$ distortions within the $\Lambda$CDM model as being ${3.6\times10^{-9}}$ and ${1.9\times10^{-8}}$, respectively. The single contribution from the dissipation of acoustic waves is $y\simeq4.1\times10^{-9}$ and $\mu\simeq2.3\times10^{-8}$, while for the adiabatic cooling of baryons one has $y\simeq-5.2\times10^{-10}$ and $\mu\simeq-3.3\times10^{-9}$.

\begin{figure}
 \centering
 \includegraphics[width=7.5 cm]{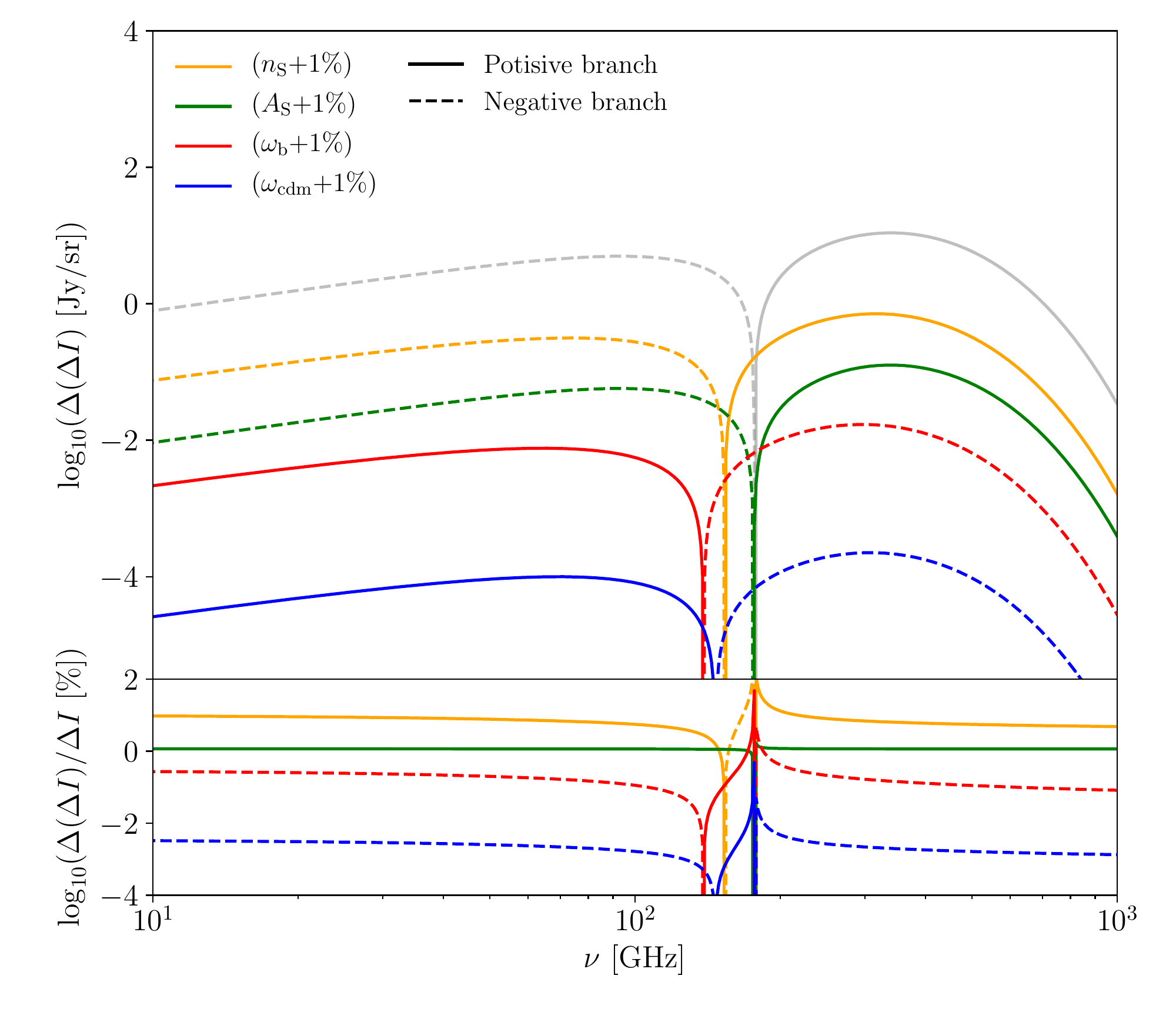}
 \includegraphics[width=7.5 cm]{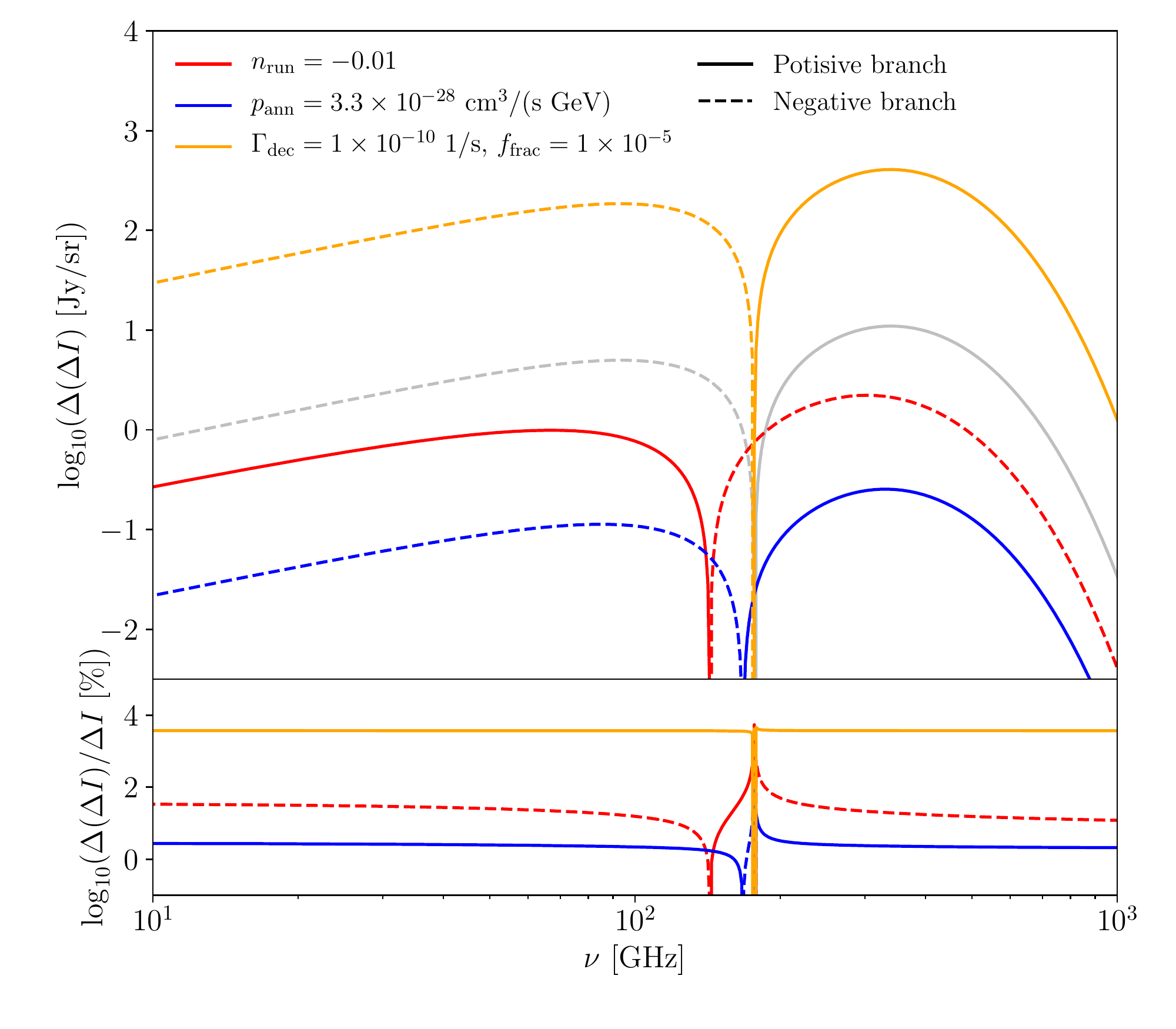}
 \caption{\textit{Left panel}: Impact of a variation in the $\Lambda$CDM parameters on the primordial SD signal. The top panel shows the difference $\Delta(\Delta I)=\Delta I_{\rm  var}-\Delta I_{\rm fid}$ (in logarithmic scale) between the total SD spectrum $\Delta I_{\rm  var}$ obtained by changing a cosmological parameter by $1\%$ while keeping all the other fixed, and the fiducial $\Lambda$CDM total distortion spectrum $\Delta I_{\rm fid}$. As a reference, we also show the $\Lambda$CDM fiducial primordial SD spectrum ($\Delta I_{\rm fid}$) in dashed gray.  The bottom panel shows the same as the top panel, but in relative units. These plots show that a $1\%$ change in $n_{\rm s}$ produces the largest SD variation, of the order of $\sim10\%$, followed by $A_{\rm s}$, which induces a $\sim 1\%$ variation. \textit{Right panel}: Same as in the left panel but for the extensions of the $\Lambda$CDM model considered within this work.}
 \label{fg_LCDM_params}
\end{figure}

Of the $\Lambda$CDM parameters listed in Equation~\eqref{eq_param_set_LCDM}, only four ($\omega_\text{b}$, $\omega_\text{cdm}$, $A_\text{s}$ and $n_\text{s}$) effectively influence the shape of the SD signal, and can thus be constrained with SDs. To show the impact of the single parameters, in the left panel of Figure~\ref{fg_LCDM_params} we vary each of them by 1\% with respect to the Planck best fits (keeping all the others fixed) and display the corresponding variation of the SD signal. The parameter that affects the total signal the most is the spectral index $n_\text{s}$, which induces
variations of the order of $\Delta (\Delta I)/\Delta I \sim 10 \Delta n_s$, as can be seen from the lower panel of the figure. This is because even a small change of this quantity strongly influences the amplitude of the power spectrum at scales much smaller than the pivot scale ($k \gg 0.05\,$Mpc$^{-1}$) and, therefore, the amount of acoustic dissipation at high redshifts. On the other hand, changing the amplitude $A_\text{s}$ by 1\% results in a variation of the signal by 1\%, as expected 
and evident from the constant green line in the lower panel of the figure. The impact of $\omega_\text{b}$ is due to its role in the definition of the damping scale $k_\text{D}$ of Equation \eqref{eq_diss_ac_waves}. A higher baryon density ensures a tighter coupling of electrons and photons, thus moving $k_\text{D}$ to higher values, i.e. to smaller scales. The impact of this effect is very mild on the SD signal, of the order of $0.1\%$. Finally, since a (minor) part of the SD signal is produced during matter domination, where the expansion of the universe is mostly determined by $\omega_\text{cdm}$, the DM energy density plays an almost negligible role in the final shape of $\Delta I$, inducing variations of the order of $0.001\%$.

The right panel of Figure~\ref{fg_LCDM_params} shows the impact on the SD signal of the running of the scalar spectral index, together with a few other extensions of the $\Lambda$CDM model considered in this paper. A running of $n_{\rm run}=-0.01$, compatible with current limits, generates a variation of the SD signal of the order of $10\%$, similar to the modifications produced by $n_{\rm s}$.

\subsection{Annihilating and decaying relic particles}
In addition to the contribution to SDs from the $\Lambda$CDM model, it is also interesting to consider exotic energy injection scenarios. As representative cases, here we consider the annihilation and the decay of relic particles. Since in these scenarios the intensity of the distortion signal directly depends on the particle physics nature of the DM, an eventual SD measurement could be very interesting from a model building perspective. When exploring these models, their effects add to the ones described for the $\Lambda$CDM case.

Firstly, we consider a self-annihilating DM particle $\chi$ and its antiparticle $\bar \chi$. We allow such particles to possibly coexist with another fully stable cold DM species, and we call $f_{\rm frac}$ the annihilating DM fraction. Adopting the parametrization of \cite{Zhang2006, galli_2009}, the energy injection rate predicted by this model takes the form
\begin{equation}\label{eq: heating DM ann}
    \dot{\mathcal{Q}}= \rho_{\mathrm{cdm}}(z)^{2} p_{\mathrm{ann}}\,,
\end{equation}
where 
\begin{equation}
    p_\text{ann}= f_{\mathrm{frac}} f_{\mathrm{eff}} \frac{\langle\sigma v\rangle}{\mathrm{M}_{\chi}}
\end{equation}
represents the annihilation efficiency and contains information on the mass $\mathrm{M}_{\chi}$ and fractional density $f_{\mathrm{frac}}$ of the particle, on the thermally averaged cross-section $\langle\sigma v\rangle$ of the annihilation process, as well as on the energy injection efficiency $f_{\mathrm{eff}}$. We adopt here an on-the-spot approximation, i.e. the energy emission and absorption are assumed to happen at the same redshift.

Similarly to the previous scenario, in this case we have a 6+1 parameter extension of the $\Lambda$CDM model, with the set of parameters:
\begin{equation}\label{eq_param_set_ann}
    \{ \omega_\text{b}, \omega_\text{cdm}, 100 \theta_s, \lnAs, n_\text{s}, \tau_\text{reio} \} + p_\text{ann}\,.
\end{equation}
Assuming the largest value allowed by Planck\footnote{The units adopted in \cite{planck_2018_cosmo_params} are cm$^3$/(s GeV). A useful conversion is $1\times 10^{-28}$ cm$^3$/(s GeV) = $6.2\times 10^{-25}$ m$^3$/(s J).} for $p_\text{ann}$, i.e. ${p_\text{ann}=3.3\times 10^{-28}}$ cm$^3$/(s GeV), and the best-fit value of the other parameters \cite{planck_2018_cosmo_params},   we obtain values for the $y$ and $\mu$ parameters in the order of $7.1\times10^{-11}$ and $5.5\times10^{-10}$, respectively.

On the other hand, the energy release rate in the case of decaying relic particles can be parametrized according to \cite{chen_kamionkowski_2004} as
\begin{equation}
 \dot{\mathcal{Q}}=\rho_{\mathrm{cdm}}(z) f_{\mathrm{frac}} f_{\mathrm{eff}} \Gamma_{\mathrm{dec}} e^{-\Gamma_{\mathrm{dec}} t}\,,
	\label{eq_DM_dec_ERR}
\end{equation}
where $f_{\rm frac}$ is the fraction of decaying DM and $\Gamma_{\rm dec}$ is the particle decay width. Due to the exponential factor, the decay process starts to be inefficient as soon as the age of the universe is comparable to the lifetime of the particle, $\tau_{\rm dec} \sim \Gamma_{\rm dec}^{-1}$.
For simplicity, we will assume that $f_{\mathrm{eff}}=1$ and only keep $f_{\rm frac}$ as a free parameter. Note that in principle the two are completely degenerate, and thus the constraint on $f_{\rm frac}$ depends on the specific decay channel considered and therefore on the value of $f_{\mathrm{eff}}\le 1$. As in the case of annihilating DM particles, we assume an on-the-spot approximation (for a possible generalization of these approximations see e.g. \cite{Poulin:2016anj,Acharya:2019owx}).

Differently than for the previous heating mechanisms, in the case of DM decay we have a 6+2 parameter extension:
\begin{equation}\label{eq_param_set_dec}
    \{ \omega_\text{b}, \omega_\text{cdm}, 100 \theta_s, \lnAs, n_\text{s}, \tau_\text{reio} \} + f_{\rm frac},\Gamma_{\rm dec}\,.
\end{equation}
For parameter values such as $f_{\rm frac}=1\times 10^{-5}$ and $\Gamma_{\rm dec}=1\times 10^{-10}$ 1/s, which are well within FIRAS bounds, one obtains $y \sim 1.2\times10^{-7}$ and $\mu \sim 9.6\times10^{-7}$. Note, however, that the relative amplitude of $y$ and $\mu$ distortions generated by DM decay strongly depends on the particle's lifetime.

The right panel of Figure~\ref{fg_LCDM_params} shows the impact on the SD signal of DM annihilation and decay. A DM annihilation process with $p_{\rm ann}=3.3\times 10^{-28}$ cm$^3$/(s GeV), compatible with current limits, generates a variation of the SD signal of the order of 2\%, while a DM decay with ${f_{\rm frac}=1\times 10^{-5}}$ and $\Gamma_{\rm dec}=1\times 10^{-10}$ 1/s produces variations of a factor of order 50, making it the most impactful process considered within this work.

\section{Experimental setups}\label{sec_exp_config}
\begin{figure}
    \centering
    \includegraphics[width=10 cm]{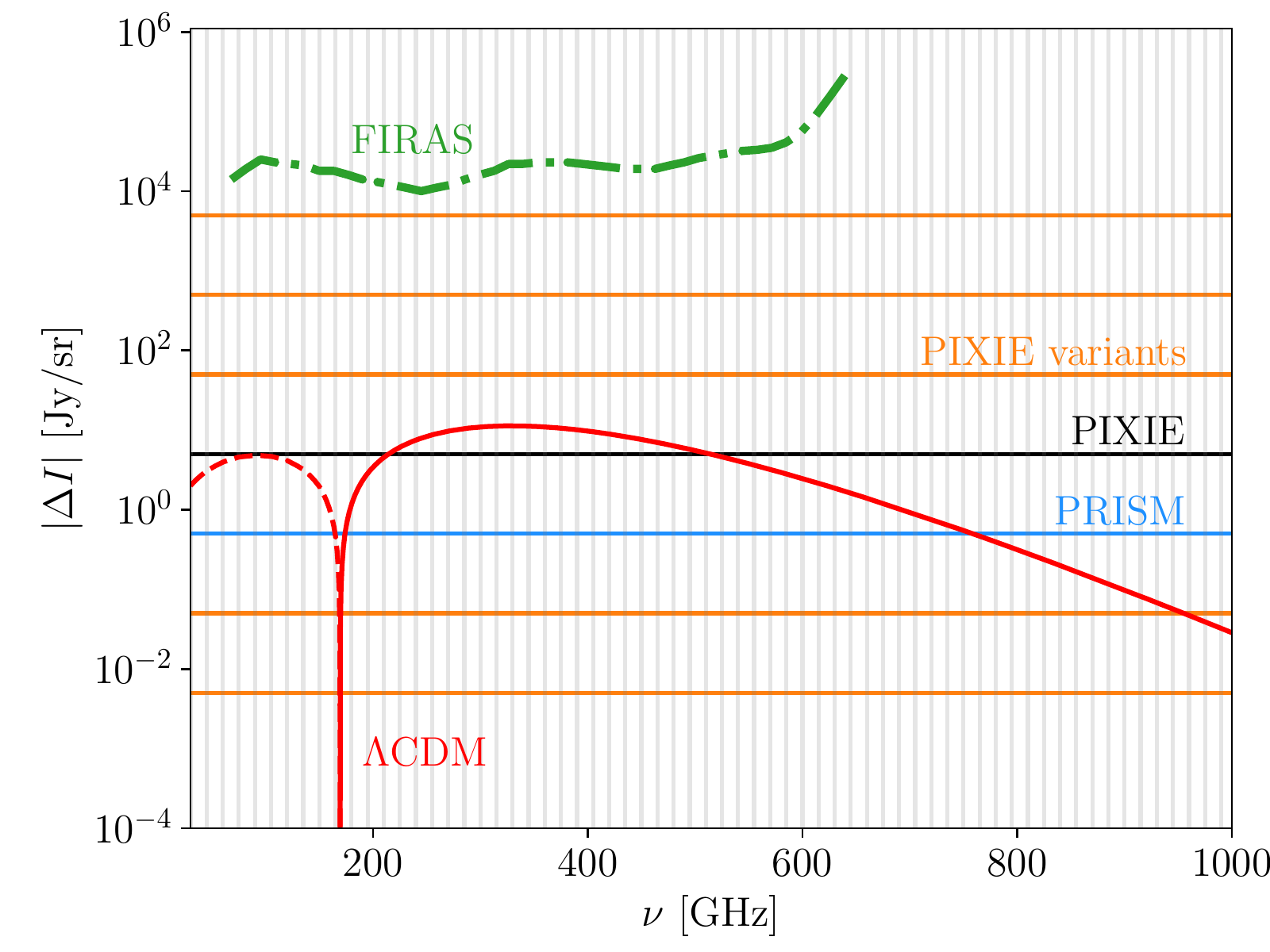}
    \caption{Total primordial SD signal expected within the $\Lambda$CDM scenario (red line) in absolute values, with positive (negative) values shown with a solid (dashed) line. The signal includes the contribution of the adiabatic cooling of baryons and small-scale acoustic wave dissipation computed according to Equation~\eqref{eq_diss_ac_waves}, but does not include foregrounds. We compare this signal to the sensitivity of the experimental setups considered within this work. For the sensitivity of FIRAS we employ the same frequency binning  as in Table 4 in \cite{fixsen_1996}. For the PIXIE variants, we use the same binning as PIXIE, indicated by the gray vertical lines. For illustrative purposes, here we refer to PIXIE10 as PRISM.}
    \label{fg_sens_pixie_var}
\end{figure} 

Within this work, we consider several experimental configurations spanning a variety of different sensitivities and possible observational environments. In this section we provide an overview of their most relevant characteristics.

\subsection{PIXIE and its variants}\label{sec_pixie}
The Primordial Inflation Explorer (PIXIE) is a mission proposed to NASA aimed at probing the nature of primordial inflation with measurements of the CMB B-modes linear polarization caused by such an inflationary epoch \cite{kogut_2011}. However, the experiment was not only designed for B-modes, but also for the observation of CMB SDs. PIXIE consists of a polarizing Fourier Transform Spectrometer (FTS) that synthesizes $\sim$ 400 frequency channels, ranging from 30~GHz to 6 THz, with a photon noise equal to ${\Delta I_\text{noise} \simeq 5 \times 10^{-26} \text{ W/m}^2 \text{/Hz/sr}}$ (more details on how this value is obtained, such as the integration time and the number of detectors assumed, can be found in \cite{chluba_jeong_2014} and Appendix E.1 of \cite{lucca_2019}). For the purposes of this work, and as also commonly done in the literature, we assume for PIXIE a constant channel frequency resolution of $\Delta \nu = 15 \text{ GHz}$ in the range ${30\text{ GHz} - 1\text{ THz}}$ and uncorrelated noise for each resulting bin (64 in total). Hereafter, we will refer to this setup as PIXIE.

Furthermore, we consider several other experimental configurations with identical frequency binning as PIXIE, but with different sensitivities ranging from 1000 times smaller to 1000 times larger noise than that of PIXIE\footnote{Specifically, three configurations with better sensitivity than PIXIE and three with worse sensitivity.}. Hereafter, we will refer to these configurations as PIXIE variants, and label the cases with better sensitivity as PIXIE10, PIXIE100 and PIXIE1000 (for configurations with 10, 100 and 1000 times improved sensitivity, respectively). In Figure \ref{fg_sens_pixie_var} we show the sensitivity of these experimental concepts, along with those of PIXIE and FIRAS, together with the primordial $\Lambda$CDM signal (which is what we will mostly focus on within this work) in red for reference.

As a remark, note that the Polarized Radiation Imaging and Spectroscopy Mission (PRISM), a mission proposed to ESA for the investigation of early universe physics \cite{prism_2014}, is predicted to reach sensitivities roughly one order of magnitude better than those of PIXIE ($\Delta I_\text{noise} \simeq 6 \times 10^{-27} \text{ W/m}^2 \text{/Hz/sr}$), and more accurate performances are reported in Table 2 of \cite{prism_2014}. Assuming, therefore, an experimental configuration with the same frequency binning as PIXIE and 10 times better sensitivity, i.e. $\Delta I_\text{noise} \simeq 5 \times 10^{-27} \text{ W/m}^2 \text{/Hz/sr}$, it is in principle possible to compare the results obtained for PIXIE10 to those within the reach of PRISM. Similarly, it is also possible to place the recently proposed Super-PIXIE mission~\cite{Kogut:2019vqh} and its upgraded Voyage 2050 version \cite{Chluba:2019nxa} in the sensitivity range between PIXIE10 and PIXIE100, although for these cases the improved low-frequency sensitivity band of the former configurations does not allow for a direct comparison.

\subsection{Experiments in different environments}\label{sec_exp_diff_env}
We also consider three additional instrumental configurations for possible future experiments, hereafter referred to as ground, (stratospheric-) balloon and satellite configurations. This analysis is aimed at testing the feasibility of an SD detection with a ground-based experiment, and at showing the level of improvement that could be achieved by further investing in a balloon-borne or satellite experiment.
    
For these three experimental configurations, we have assumed that the instrument measures the signal laying within two frequency ranges centered at 150 GHz and 220 GHz, each covering a bandwidth corresponding to 25\% of the central value, and with realistic optics and atmosphere emission. The particular choice made for the frequency intervals is dictated by the favorable atmospheric windows for the ground-based configuration (see e.g. \cite{Battistelli2012} and references therein) and applied also to the other setups to ease the comparison between environments. In other words, for the balloon and satellite configurations we use the same frequency settings to isolate the impact of just improving the sensitivity of the experiment on the considered parameters. We assume that the experiments are limited by the radiation photon noise on the detectors, which is in turn dominated by the atmosphere for the ground-based experiment, by the residual atmosphere and by the optics emission for the balloon, and by the CMB for the satellite configuration. The noise equivalent power (NEP) for these configurations has been calculated taking into account the photon noise from the optics and the atmosphere emission only, although for ground-based operations one should also account for the fluctuations of the atmosphere (see e.g. \cite{Errard:2015twg}). Concerning the photon noise from foregrounds, in particular galactic dust, we estimate that the contribution is negligible for the considered configurations (with contributions up to an order of $5\times 10^{-20}-5\times 10^{-18}$ W/Hz$^{1/2}$ for both frequency windows), especially considering that these experiments would observe patches of the~sky with low galactic emission.
    
Photon noise levels have been derived considering the three different experimental setups. For the satellite configuration we assumed a cold optics configuration and calculated the photon noise level arising from the CMB itself including both photon (shot) noise and the bunching component (see e.g. \cite{Lamarre1995}). For the ground-based configuration, we accounted for the atmospheric emission using \textit{am}\footnote{\href{https://www.cfa.harvard.edu/~spaine/am/}{\textit{am:} Atmospheric Model, Website}} \cite{scott_2019_am}, a radiative transfer tool developed for microwave to sub-millimeter wavelengths atmospheric emission, which accounts for realistic antarctic conditions \cite{Battistelli2012}. In the balloon-borne configuration, we have accounted for both the residual atmospheric emission (see e.g. \cite{Masi:2019wqa}) and the optics emission. For the latter, we have assumed mirrors, lenses and a window with 1\% emissivity. 
\begin{table}
    \begin{center}
     \begin{tabular}{|c |c |c |c |c|}
         \hline\rule{0pt}{2.0ex} 
	        & $\nu \text{ [GHz]}$ & $\Delta \nu \text{ [GHz]}$ & NEP  [W/Hz$^{1/2}]$ & $\Delta I_\text{noise} \text{ [W/m}^2 \text{/Hz/sr]}$ \\
		    \hline\rule{0pt}{2.0ex} 
      PIXIE & 30 - 1005 & 15 & $7.0 \times 10^{-17}$ & $5.0 \times 10^{-26}$\\
      \hline\rule{0pt}{2.0ex} 
   SATELLITE & 135 - 165 & 5 & $1.0 \times 10^{-17}$ & $3.8 \times 10^{-26}$\\
		    & 195 - 245 & 5 & $1.0 \times 10^{-18}$ & $8.4 \times 10^{-27}$\\
		    \hline\rule{0pt}{2.0ex} 
 	    BALLOON & 135 - 165 & 5 & $2.0 \times 10^{-17}$ & $7.6 \times 10^{-26}$\\
      & 195 - 245 & 5 & $2.5 \times 10^{-17}$ & $2.1 \times 10^{-25}$\\
      \hline\rule{0pt}{2.0ex} 
   GROUND & 135 - 165 & 5 & $6.5 \times 10^{-17}$ & $2.5 \times 10^{-25}$\\
		    & 195 - 245 & 5 & $1.0 \times 10^{-16}$ & $8.4 \times 10^{-25}$ \\
 	    \hline
  \end{tabular}
 \end{center}
    \caption{Instrumental specifications of PIXIE \cite{kogut_2011}, and  three other possible future experimental configurations centered in two frequency bands. The second column represents the frequency range explored by each experiment, while the third column represents the bandwidth of each frequency channel matching the atmospheric windows. The fourth column represents the corresponding detector's noise equivalent power (NEP) and the last column the final sensitivity. In this paper, we also consider experiments with NEP (and thus sensitivities) 10, 100 or 1000 better or worse than PIXIE.} \label{tb_photon_noise}
\end{table}
\begin{figure}
	\centering
	\includegraphics[width=10 cm]{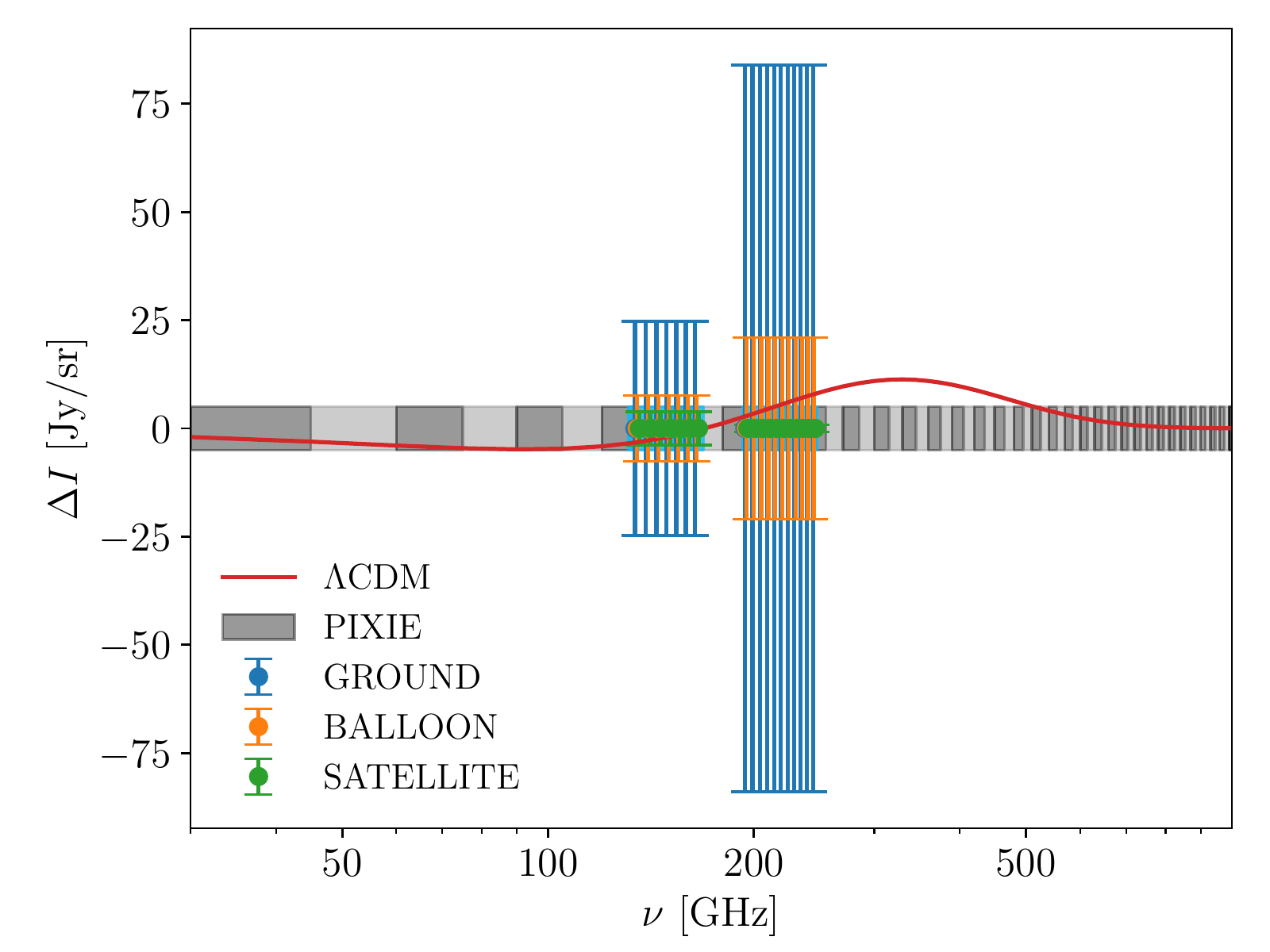}
	\caption{Final sensitivity for different experimental configurations (in units of  $\mathrm{Jy}/\mathrm{sr} =10^{-26} \mathrm{W}/\mathrm{m}/\mathrm{sr}/\mathrm{Hz^{-1}}$) . The gray area represents the sensitivity of PIXIE. The blue, orange and green error bars represent respectively the sensitivity of the configurations of an experiment from the ground, balloon and satellite, all with seven channels, 5 GHz distant, centered in two frequency bands (150 GHz and 220 GHz) with 25\% width each. The error bars representing the balloon experiment noise are plotted in the real value of the frequency, while for the satellite and ground cases there is an offset of $\pm 1 \text{ GHz}$ for the sake of graphical clearness. The red line represents the predicted $\Lambda$CDM distortion signal assuming Planck bestfits.}
	\label{fg_sensitivities}
\end{figure}

The final detector sensitivity for each single frequency channel has been derived and calculated from the NEP according to Equation (3.3) in \cite{kogut_2011},
\begin{equation}
    \Delta I_\text{noise} = \frac{\text{NEP} / \sqrt{\tau / 2}}{A \Omega \Delta \nu (\alpha \epsilon f)}\,,
\end{equation}
where $\tau$ is the integration time, $\Delta \nu$ the frequency bandwidth, $\alpha$ its absorptivity, $\epsilon$ the source emissivity and $f$ the transmissivity of the optics. In our calculation, we have assumed a detector with ${\alpha f = 0.4}$, ${\epsilon = 1}$, twenty modes of the radiation and integrated over 1 year mission time. We have assumed a diffraction limited detector model, i.e. the \textit{\'etendue} ${A \Omega = N \lambda^2}$ being constant in each observed band, where $N$ is the radiation modes and $\lambda$ the smallest wavelength in each band, corresponding to 0.18 cm and 0.12 cm respectively. In Table \ref{tb_photon_noise} we report the value of the different photon noises within each frequency band. These are also illustrated in Figure \ref{fg_sensitivities}. From this figure it is already possible to notice that neither a ground-based nor a balloon-borne experiment will be able to detect the minimal primordial distortion expected in the $\Lambda$CDM model. A PRISM-like experiment will be barely able to do so, as also confirmed by the results reported in Section \ref{sec_res}.

\section{Methodology}\label{sec_method}
In the previous sections we have introduced some well-studied representative examples of models causing an energy injection in the CMB photon field.

For the numerical evaluation of the many cosmological quantities involved, such as the CMB power spectra, we use the Boltzmann solver \textsc{class}\footnote{\href{http://class-code.net/}{CLASS: the Cosmic Linear Anisotropy Solving System, Website}} \cite{lesgourgues_class_i, lesgourgues_class_ii}. In particular, we use the latest version of the code which allows the calculation of energy injection rates and SD spectra, as described in \cite{lucca_2019}, that will be released very soon as v3.0.
Furthermore, to forecast the constraints on cosmological parameters we use the MCMC sampler \textsc{MontePython}\footnote{\href{https://github.com/brinckmann/montepython_public}{MontePython 3: Website}} \cite{audren_2013, brinckmann_2019}. The implementation of the specific likelihoods required for SD experiments are also described in \cite{lucca_2019}. In the following subsections, we will report some of the details related to the implementation of the SD modelling and forecasting in \textsc{class} and \textsc{MontePython}.
	
\subsection{The \textsc{class} implementation}\label{sec_class}
In full generality, the SD signal can be parametrized as
\begin{equation}
    \Delta I^\text{tot} = \Delta I^\text{T} + \Delta I^y + \Delta I^\mu + \Delta I^\text{R} + \Delta I^\text{reio} + \Delta I^\text{fg}\,,
\end{equation}
where $\Delta I^\text{T}$ represents temperature shifts, $\Delta I^y$, $\Delta I^\mu$ and $\Delta I^\text{R}$ the Compton $y$, the chemical potential $\mu$ and the residual distortions, $\Delta I^\text{reio}$ the contribution from late-time SD  sources (such as cosmic reionization, the ICM or IGM), and $\Delta I^\text{fg}$ all foreground contaminations.

According to this formalism, the evaluation of the SD signal is performed using \textsc{class}, which calculates the heating history and the distortion signal for a specific energy release scenario and experimental setting, making use of the Green's function approximation introduced in \cite{chluba_greens} for fast computations. In this approximation, once the cosmological model has been defined, the SD intensity can be linearized as
\begin{equation}
\label{eq:deltaI}
    \Delta I(\nu,z_0) = \int_{z_0}^{\infty} G_\text{th} (\nu,z') \frac{\mathrm dQ(z')/dz'}{\mathrm \rho_\gamma(z')} \mathrm dz'\,,
\end{equation}
where $\rho_\gamma (z)$ is the energy density of the CMB photons and
\begin{equation}
\label{eq:shapes}
    G_{\mathrm{th}}\left(\nu, z^{\prime}\right)=\mathcal{G}(\nu) \mathcal{J}_{g}\left(z^{\prime}\right)+\mathcal{Y}(\nu) \mathcal{J}_{y}\left(z^{\prime}\right)+\mathcal{M}(\nu) \mathcal{J}_{\mu}\left(z^{\prime}\right)+R\left(\nu, z^{\prime}\right)
\end{equation}

is the Green's function of the SD signal. Here, $\mathcal{G}$, $\mathcal{Y}$ and $\mathcal{M}$ are the temperature-shift, the $y$-type and the $\mu$-type spectral shapes, respectively, and $\mathcal{J}_{g}$, $\mathcal{J}_{y}$ and $\mathcal{J}_{\mu}$ are the corresponding branching ratios. Finally, $R\left(\nu, z^{\prime}\right)$ represents the residual distortions which are not captured by the other ones. This decomposition of $\Delta I$ in Green's function and heating rate presents the important advantage of shifting the whole model dependence of the final signal into the shape of $Q$, while $G_\text{th}$ is completely model independent. Additional details on the assumptions on which this expansion is based are provided e.g. in \cite{chluba_greens, chluba_jeong_2014, Chluba:2015hma} and \cite{lucca_2019} (see Section 3.2.1 therein).

For the calculation of the different heating rates, we follow the prescriptions described in Section \ref{sec_scenarios}. In our analysis we included only the dissipation of acoustic waves and the adiabatic cooling of baryons and electrons as the $\Lambda$CDM prediction. The contributions from the CMB dipole  and from the cosmic recombination radiation (CRR) have been neglected. Note, however, that both of these effects are predicted to have only a minor impact on the final result (see e.g. Figure 1 of \cite{chluba_2016}). Dedicated forecasts focused on the role of the CRR can be found in \cite{Hart:2020voa}.

Moreover, we rely on the publicly available Green's functions of the CosmoTherm repository\footnote{\href{http://www.jb.man.ac.uk/~jchluba/Science/CosmoTherm/Welcome.html}{CosmoTherm: Website}}, computed following  the method first introduced in \cite{chluba_greens} and distributed together with \textsc{class} 3.0.
There, all frequency-dependent components of $G_\text{th}$ are discretized. Therefore, the spectral shapes $\mathcal{G}$, $\mathcal{Y}$, $\mathcal{M}$ and $R$ are fixed vectors in frequency space, with frequency range and resolution depending on the considered experiment. To determine the branching ratios $\mathcal{J}_{g}$, $\mathcal{J}_{y}$ and $\mathcal{J}_{\mu}$, the total Green's function $G_\text{th}$ is then projected onto these vectors. However, since these vectors are not orthogonal, the projection is not unambiguous\footnote{Even beyond frequency coverage, channel spacing and sensitivity there exists an \textit{intrinsic} degeneracy within the Gram-Schmidt orthogonalization process (which is employed in \cite{chluba_greens}). Consider, for example, the vectors $\vec{v}_1=(1,1)$ and $\vec{v}_2=(1,-2)$. Then, depending on the order of the Gram-Schmidt process we obtain (up to normalization) $\vec{u}_1=(1,1)$ and $\vec{u}_2=(1,-1)$ or $\vec{u}_2'=(1,-2)$ and $\vec{u}_1'=(2,1)$, which are not equivalent, and would give different amplitudes for the components $1$ and $2$. Additionally, note also that there are alternatives beyond the Gram-Schmidt orthogonoalization employed in \cite{chluba_greens}.} (see Section 3.2.2 of \cite{lucca_2019}), i.e. the branching ratios depend on the choice of the projection procedure, as well as on the frequency range and resolution of the experiment considered.

In this formalism, we define the experimentally determined $\mu$ and $y$ parameters as
\begin{equation}
    \label{eq_def_mu_y}
    \mu = \int_{z_0}^{\infty} \mathcal{J}_{\mu}\left(z^{\prime}\right) \frac{\mathrm dQ(z')/ \mathrm dz'}{\mathrm \rho_\gamma(z')} \mathrm dz' \quad \text{and} \quad y = \int_{z_0}^{\infty} \mathcal{J}_{y}\left(z^{\prime}\right) \frac{\mathrm dQ(z')/ \mathrm dz'}{\mathrm \rho_\gamma(z')} \mathrm dz'\,.
\end{equation}
Note that in this definition, $\mu$ and $y$ correspond to the amplitudes of the spectral shapes $\mathcal{M}$ and $\mathcal{Y}$, respectively, that one obtains from a least-square fit of the total $\Delta I(x)$ in Eq.~\eqref{eq:deltaI} for given a frequency range and resolution (encoded by the branching ratios $\mathcal{J}$). These amplitudes can thus slightly differ from the $\mu$ and $y$ quantities one can theoretically define (see e.g. \cite{hu_silk_1993}). However, the two should agree in the limit of infinite frequency range and resolution.\footnote{Since the the branching ratios $\mathcal{J}_{\mu}$ and $\mathcal{J}_{y}$ depend on frequency resolution and range of a given SD experiment as well as on the projection procedure, the $\mu$ and $y$ parameters inferred from Equation \eqref{eq_def_mu_y} can vary according to these possible choices. Instead, one could choose to display the true physical $y$ and $\mu$ parameters, which would correspond also to those measured by an experiment with infinitely precise frequency coverage. However, this would not be a perfectly fair comparison, as a given experiment's bounds will depend on the finite precision of its frequency resolution and the corresponding fitting of the SD amplitudes. This is consistent with the idea that a given experiment will not be able to perfectly differentiate between the different contributions to the total signal, such as the $y$, $\mu$, or residual distortion, which are usually fitted to the total signal as in the case of FIRAS~\cite{fixsen_1996}.}
Since the heating rate $\mathrm dQ(z)/\mathrm dz$ depends on the cosmological parameters as described in Section~\ref{sec_scenarios}, these measured $\mu$ and $y$ parameters are then derived quantities. Their uncertainties are thus determined by the combined power of the experiments used to constrain the cosmological parameters of the considered model, which in this paper are CMB anisotropy and SD experiments. Therefore, the uncertainties on the $\mu$ and $y$ parameters reported in Section \ref{sec_res} do not reflect the constraining power of a given SD experiment alone on these spectral shapes (see Section~\ref{sec_lcdm_params} for additional details), which are however commonly available in the original papers such as \cite{fixsen_1996} for FIRAS and \cite{kogut_2011} for PIXIE.
\begin{figure}
 \centering
 \includegraphics[width=7.6 cm]{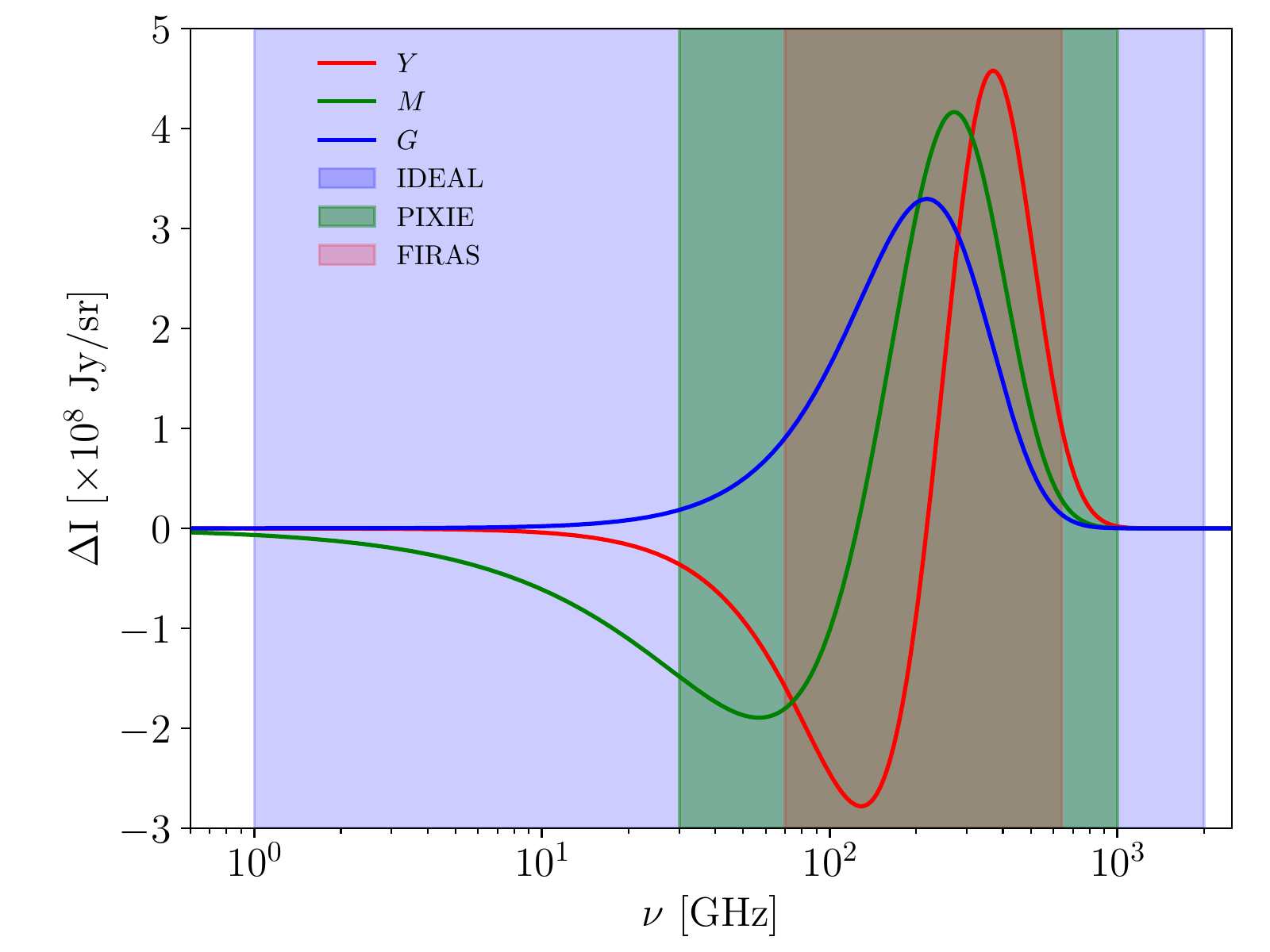}
 \includegraphics[width=7.6 cm]{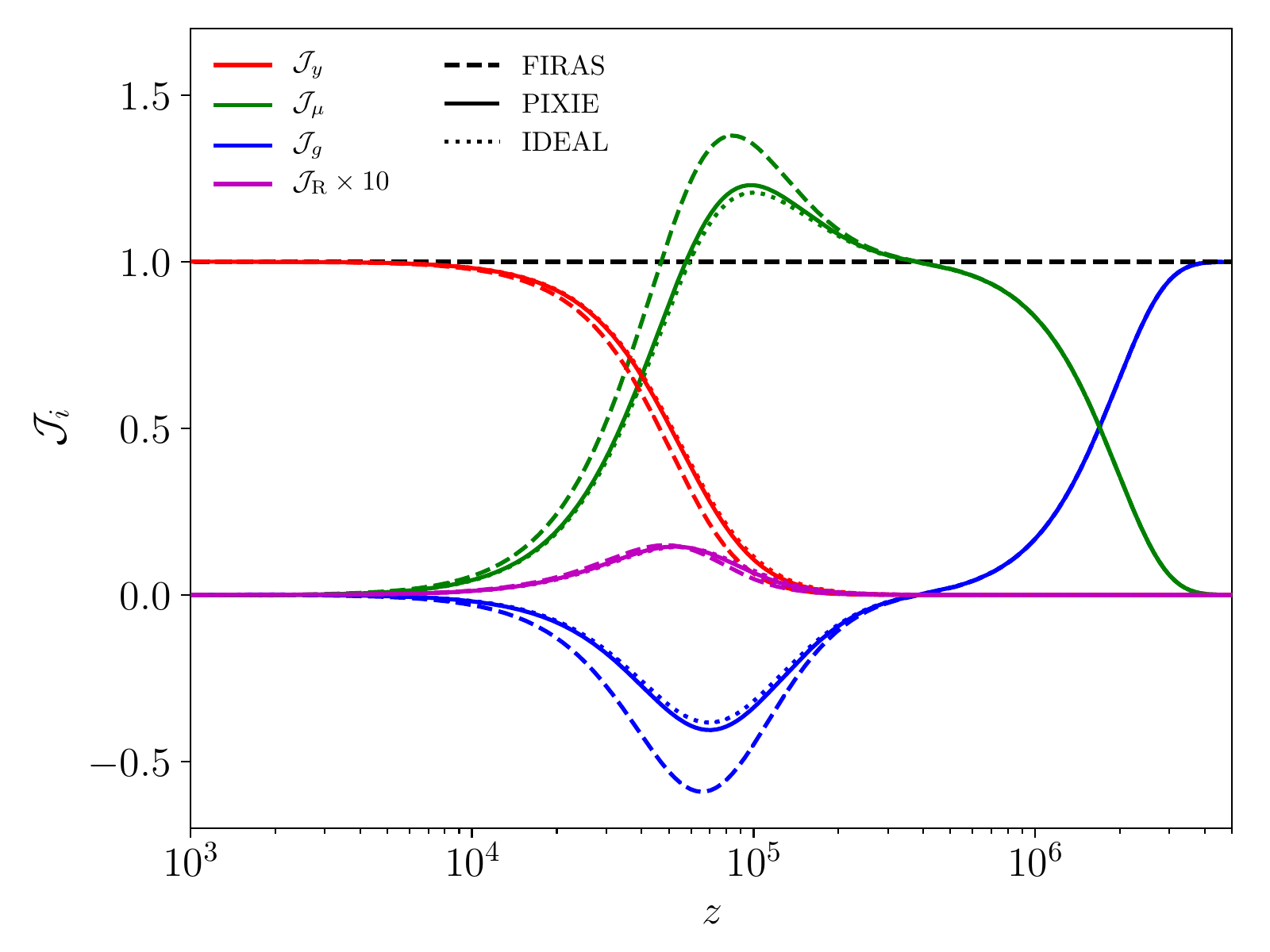}
    \caption{\textit{Left panel}: SD shapes assuming an instantaneous energy injection during the $y$-era (red line), during the $\mu$-era (green line) and at early times (blue line), same as in Figure 1 of \cite{lucca_2019}. The vertical red, green and blue bands represent the frequency range covered by FIRAS, PIXIE and an ideal detector with improved characteristics referred to as IDEAL. \textit{Right panel}: Comparison between the branching ratios of the different SD components for the three aforementioned configurations.}
 \label{fg_br}
\end{figure}

To illustrate this point, we compare three experimental configurations featuring different frequency ranges and resolutions: FIRAS and PIXIE (defined according to Section \ref{sec_exp_config}), as well as an ``ideal'' detector with much wider frequency range and higher resolution with respect to the other two, i.e. $\{\nu_{\rm min}, \nu_{\rm max}, \Delta \nu \} = \{ 1 \text{GHz}, 2 \text{THz}, 1 \text{GHz}\}$, which will be referred to as IDEAL. The left panel of Figure \ref{fg_br} shows the $\mathcal{G}$, $\mathcal{Y}$ and $\mathcal{M}$ spectral shapes, together with the range of frequencies probed by each of the three configurations. The figure shows that IDEAL covers the whole frequency space where the shapes are non-zero, while both PIXIE and FIRAS are insensitive to the low-frequency tail of the curves. As a consequence, the branching ratios are different for each experimental configuration, as shown in the right panel of Figure~\ref{fg_br}. There, the branching ratio of the $\mu$ distortion is enhanced for FIRAS with respect to PIXIE, while the $y$ branching ratio is reduced. A consequence of this is clearly visible, for instance, in the right panel of Figure 7 of \cite{lucca_2019}, where the FIRAS contour is shifted towards lower values of $y$ and higher values of $\mu$ with respect to PIXIE-like configurations (this effect is not noticeable in Figure \ref{fg_contour_run_y_mu} because of the marginalization over late-time SD sources performed in this work)\footnote{We re-iterate that this is a result of considering the $y$ and $\mu$ parameters a given experiment would measure, not those generically predicted by the given model.}. Interestingly, although the difference of coverage between PIXIE and IDEAL in the low energy tail is substantial, the deviation of PIXIE from the ideal setup is minor, suggesting that PIXIE-like frequency arrays are already close to optimal for the purpose of disambiguation of the $\mu$ and $y$ distortion shapes, and considering the limited foreground model described in Section \ref{sec_montepython}.

In any case, the dependence of the amplitude of the $\mu$ and $y$ parameters on the experimental configuration is not a problem, since the final observable is the total SD spectrum, which is the weighted sum of all shapes and which is thus independent of the details of the projection. We base our results solely on this total SD spectrum. However, as already pointed out in Section 2.2.1 and Figure 3 of \cite{chluba_jeong_2014} as well as in Section 4.1 of \cite{lucca_2019}, this means that the same heating rate history can result in different measured distortion amplitudes\footnote{Of course, a given experiment might attempt to correct for the known biases on the measured $y$ and $\mu$ parameters in order to infer the underlying physical $y$ and $\mu$ parameters. Nevertheless, since in this work we focus solely on the constraints on the involved cosmological parameters, we leave such an analysis for future work.} depending on the experiment and projection choices considered.

\subsection{The \textsc{MontePython} implementation} \label{sec_montepython}
\begin{table}
   	\begin{center}
   		\begin{tabular}{ | c | c | c | c | c | c | }
       		\hline\rule{0pt}{2.0ex}
    	    $\omega_\text{b}$ & $\omega_\text{cdm}$ & $100 \theta_s$ & $\lnAs$ & $n_\text{s}$ & $\tau_\text{reio}$ \\
       		\hline\rule{0pt}{2.0ex}
       		0.022377 & 0.1201 & 1.0411 & 3.0447 & 0.9659 & 0.0543 \\
       		\hline
   		\end{tabular}
   	\end{center}
   	\caption{Values of the cosmological parameters used for the fiducial model \cite{planck_2018_cosmo_params}, which assumes a $\Lambda$CDM scenario.}
    \label{tb_fiducials}
\end{table}

In order to constrain the cosmological parameters involved in each model considered within this work, we perform a number of MCMC scans using the \textsc{MontePython} code \cite{brinckmann_2019}.

We produce synthetic mock data for all of the CMB anisotropy and SD experiments considered, using as a fiducial the $\Lambda$CDM model with best-fit cosmological parameters from the Planck mission \cite{planck_2018_cosmo_params} reported explicitly in Table \ref{tb_fiducials}. We use simulated data, rather than real data, also when considering completed missions such as Planck or FIRAS. This ensures that all of the experiments we include in our forecasts share exactly the same fiducial model. The experimental specifications used to simulate CMB anisotropy experiments such as Planck, \lb\ \cite{Suzuki2018LiteBIRD} and \sfour\ \cite{Abazajian:2016yjj, CMB_S4} are detailed in Section~3 and Table~5 of \cite{Brinckmann:2018owf}. The specifications used for SD experiments such as FIRAS and PIXIE are detailed in Section 3.3 of \cite{lucca_2019}. To discuss the detection capability of variants of the PIXIE experiment, as detailed in Section~\ref{sec_pixie}, or of ground-based, balloon-borne and satellite-based future experiments, as described in Section~\ref{sec_exp_diff_env}, we implement likelihoods analogously to those of FIRAS and PIXIE.

For each energy release scenario introduced in Section \ref{sec_scenarios}, we then forecast the constraints on the relevant cosmological parameters that different combinations of CMB anisotropy and SD experiments will be able to achieve, with particular regard to their detectability and corresponding uncertainties. In particular, we perform forecasts for the constraints on the 6 $\Lambda$CDM parameters, the running of the scalar spectral index and the parameters describing the DM annihilation or decay scenario.

In forecasting the capability of an SD experimental concept, we do not include the presence of foreground contamination such as galactic thermal dust, cosmic infrared background, synchrotron, free-free, integrated CO, and anomalous microwave emission (see e.g. \cite{abitbol_2017} for a detailed overview of foreground effects on SD detection). These can deteriorate the sensitivity of the overall signal by a factor of order 10 -- depending on the priors on the foreground parameters (see e.g. Tables 3 and 4 of \cite{abitbol_2017} for a quantitative evaluation). 

However, we do account for the contribution from the SZ effect in the ICM, which is the dominant late-time source of distortions due to its much higher temperature with respect to reionization and the IGM. This mainly contributes to the $y$ signal, and it is therefore strongly degenerate with early-universe phenomena which also produce $y$ distortions. We also marginalize over the experimental uncertainty $\Delta T$ on the monopole temperature $T_0$, as explained in Section 3.3 of \cite{lucca_2019}. This deteriorates the precision of the final SD signal because of the degeneracies that the temperature shifts included in $\Delta I^T$ share with the $y$ and $\mu$ distortions.

To marginalize over the contribution from the ICM, we use the implementation of the thermal (relativistic and non-relativistic) and kinematic SZ effects (kSZ) \cite{Zeldovich_1969inter} already present in \textsc{class} and based on \cite{chluba_2012fast} (see Equations (2.54)-(2.58) of \cite{lucca_2019} and references therein for more details on the \textsc{class} implementation). Since this effect is predominantly proportional to the optical depth of the electron plasma $\Delta\tau$ and the electron temperature $T_e$ \cite{1970Ap&SS...7....3S,chluba_2012fast, Chluba:2012py, Hill:2014pyr,hill_2015, Erler:2017dok, Remazeilles:2018laq}, to marginalize over late-time sources we leave $\Delta\tau$ and $T_e$ free to vary, while keeping other quantities related to the SZ effect fixed, such as the average cluster velocity with respect to the line of sight $\langle\beta\rangle$. In this parametrization $\Delta\tau$ is the mean optical depth of the ICM, which is different from $\tau_{\rm reio}$, the optical depth of reionization. For the temperature of the electron plasma we adopt the fiducial value of $T_e=1.3~{\rm keV}$, while the optical depth is set to provide a value of $y\sim 10^{-6}$ consistent with~\cite{hill_2015}.

Furthermore, importantly, out of the aforementioned contributions to the SZ effect the kSZ effect is expected to be strongly suppressed when averaged over the full sky due to the different peculiar velocities of the various clusters. However, to account for the impact of a possibly incomplete sky coverage, we conservatively set $\beta$ equal to the dispersion of the cluster velocity distribution described in \cite{Peel:2005ac}, $\langle\beta\rangle=1/300$. In fact, since  the kSZ corresponds at leading order to a temperature shift with amplitude proportional to $\langle\beta\rangle \Delta\tau$, it is degenerate with the temperature shift parameter $\Delta T$, degrading the constraints of the latter when $\langle\beta\rangle$ is non-zero. However, this degeneracy can be partially lifted by the kSZ higher order contributions, which can be captured by the setups discussed in Section \ref{sec_pixie}. We find using the aforementioned assumptions that ${\Delta T/T_0 \simeq 3\times10^{-6}}$, $5\times10^{-8}$, $5\times10^{-9}$, $6\times10^{-10}$ and $1\times10^{-10}$ for FIRAS, PIXIE, PIXIE10, PIXIE100 and PIXIE1000, respectively, almost independently from the cosmological model assumed. For the case of PIXIE1000, for which the impact of the marginalization is the strongest, taking the more realistic value of $\langle\beta\rangle=0$ reduces the uncertainty on $\Delta T$ by a factor of roughly 10, while it has a negligible impact on cosmological parameters.

For the experimental setups considered in Section \ref{sec_exp_diff_env}, we find that the limited frequency range of those experimental settings would not allow us to sufficiently break the degeneracy between $\Delta T$ and $\Delta \tau$. Therefore, in this case we can not consider the impact of this degeneracy and simply choose to set the sky-average velocity to its expected value $\langle\beta\rangle=0$, which yields $\Delta T/T_0 \simeq 1 \times10^{-7}$, $2.7 \times10^{-8}$ and $5.8 \times10^{-9}$ for the ground-based, balloon and satellite missions, respectively.

Finally, we remark again that these sensitivities to $\Delta T$ are conservative estimates and dependent on our choice of the fiducial values. A dedicated analysis with more focus on the nuisance parameters (including also $\Delta \tau$ and $T_e$) is left for future work, while here we mainly concentrate only on the cosmological parameters.

\section{Results}\label{sec_res}
In this section, we apply the numerical framework presented in Section~\ref{sec_method} to the different energy release scenarios described in Section~\ref{sec_scenarios},  and we discuss the most relevant results.

\subsection{$\Lambda$CDM}\label{sec_lcdm_params}
First of all, we focus our attention on the standard $\Lambda$CDM model and the consequences of including SDs together with completed or upcoming CMB anisotropies experiments. 

It is already known that current proposed missions such as PIXIE are not sensitive enough to improve constraints on the vanilla $\Lambda$CDM model \cite{abitbol_2017}\footnote{However, we remind the reader that PIXIE will be able to detect the SD signal from late-time sources~\cite{kogut_2011}, which we do not discuss in this Section.} -- although, obviously, they offer an extraordinary test of the model over scales that are completely different from the ones currently probed. We thus tested whether a more sensitive mission (in absence of foregrounds, which are however included in \cite{abitbol_2017}) could have a larger impact. The results are reported in the upper section of Table~\ref{tb_params_err}. First, we find that for an SD experiment such as PIXIE10 (i.e. comparable to PRISM) still no sizable improvement in the bounds on the cosmological parameters is present with respect to Planck alone. However, a futuristic experiment such as PIXIE1000 combined with Planck could, in principle, improve the bounds on $A_\text{s}$ and $n_\text{s}$ by a factor of $\sim 1.2$ and $\sim 4$, respectively, with respect to Planck alone. We also find that adding PIXIE1000 to \lb+\sfour\ would improve the constraint on $n_\text{s}$ by roughly a factor of $4$ with respect to \lb+\sfour\ alone. These are the most affected $\Lambda$CDM parameters, since they are the ones with the largest impact on SDs through the effect of anisotropy dissipation, as discussed above. As a comparison, upcoming CMB anisotropy experiments alone, such as \lb\ combined with \sfour, are expected to improve the constraints on these parameters by a factor of $\sim 2$ with respect to Planck alone. We further find that the improvement in $A_\text{s}$ from an SD experiment would also lift the known degeneracy between $A_\text{s}$ and $\tau_{\rm reio}$, improving the constraint on the latter by roughly a factor of 1.4.
\begin{table}
\begin{center}
	\fontsize{10.}{10.}{ \selectfont
	\begin{tabular}{ | c | c | c | c | c | c | c |}
   		\hline 
    	\multicolumn{7}{|c|}{$\Lambda$CDM} \\
   		\hline\rule{0pt}{2.0ex}
	     &Planck  & +FIRAS & +PIXIE & +PIXIE10 & +PIXIE100 & +PIXIE1000 \\
   		\hline\rule{0pt}{2.0ex}
   		$\sigma(100\omega_\text{b})$ & 0.015 & 0.017 & 0.016 & 0.015 & 0.015 & 0.015 \\
   		$\sigma( \omega_\text{cdm} )$ & 0.0013 & 0.0012 & 0.0013 & 0.0012 & 0.00089 & 0.00090 \\
   		$\sigma(100 \theta_s)$ & 0.00035 & 0.00032 & 0.00033 & 0.00035 & 0.00033 & 0.00034 \\
   		$\sigma (\tau_\text{reio})$ & 0.0046 & 0.0046 & 0.0044 & 0.0044 & 0.0039 & 0.0034 \\
   		$\sigma (\lnAs)$ & 0.0086 & 0.0086 & 0.0087 & 0.0085 & 0.0083 & 0.0074 \\
   		$\sigma (n_\text{s})$ & 0.0039 & 0.0037 & 0.0039 & 0.0034 & 0.0013 & 0.00092 \\
   		$\sigma (10^9 y)$ & - & 1461 & 1.0 & 0.081 & 0.010 & 0.0019 \\
   		$\sigma (10^8 \mu)$ & - & 0.073 & 0.075 & 0.065 & 0.015 & 0.0015 \\
   		\hline\rule{0pt}{2.0ex}
   		&LB+\sfour&&&&&\\
   		\hline
   		$\sigma(100\omega_\text{b})$ & 0.0034 & - & 0.0030 & 0.0033 & 0.0030 & 0.0026 \\
   		$\sigma( \omega_\text{cdm} )$ & 0.00027 & - & 0.00026 & 0.00028 & 0.00023 & 0.00020 \\
   		$\sigma(100 \theta_s)$ & 0.000086 & - & 0.000087 & 0.000087 & 0.000080 & 0.000084 \\
   		$\sigma (\tau_\text{reio})$ & 0.0019 & - & 0.0020 & 0.0021 & 0.0017 & 0.0015 \\
   		$\sigma (\lnAs)$ & 0.0033 & - & 0.0035 & 0.0037 & 0.0032 & 0.0031 \\
   		$\sigma (n_\text{s})$ & 0.0016 & - & 0.0014 & 0.0014 & 0.00074 & 0.00038 \\
   		$\sigma (10^9 y)$ & - & - & 0.92 & 0.080 & 0.0097 & 0.0012 \\
   		$\sigma (10^8 \mu)$ & - & - & 0.028 & 0.029 & 0.013 & 0.0015 \\
   		\hline
\end{tabular}
\begin{tabular}{ | c | c | c | c | c | c | c |}
        \hline
		\multicolumn{7}{|c|}{$\Lambda$CDM + running of the spectral index} \\ 
   		\hline\rule{0pt}{2.0ex}
	     & Planck & +FIRAS & +PIXIE & +PIXIE10 & +PIXIE100 & +PIXIE1000 \\
   		\hline\rule{0pt}{2.0ex}
   		$\sigma(100\omega_\text{b})$ & 0.016 & 0.016 & 0.016 & 0.016 & 0.015 & 0.015 \\
   		$\sigma( \omega_\text{cdm} )$ & 0.0013 & 0.0013 & 0.0013 & 0.0013 & 0.0013 & 0.0011 \\
   		$\sigma(100 \theta_s)$ & 0.00035 & 0.00034 & 0.00034 & 0.00034 & 0.00033 & 0.00033 \\
   		$\sigma (\tau_\text{reio})$ & 0.0049 & 0.0045 & 0.0047 & 0.0047 & 0.0044 & 0.0035 \\
   		$\sigma (\lnAs)$ & 0.0092 & 0.0094 & 0.0095 & 0.0089 & 0.0087 & 0.0077 \\
   		$\sigma (n_\text{s})$ & 0.0038 & 0.0039 & 0.0038 & 0.0038 & 0.0035 & 0.0027 \\
   		$\sigma (10^3 n_\text{run})$ & 6.5 & 6.4 & 5.9 & 2.0 & 0.85 & 0.52 \\
   		$\sigma (10^9 y)$ & - & 1509 & 0.86 & 0.11 & 0.011 & 0.0023 \\
   		$\sigma (10^8 \mu)$ & - & 0.53 & 0.47 & 0.15 & 0.014 & 0.0016 \\
   		\hline\rule{0pt}{2.0ex}
   		&LB+\sfour&&&&&\\
   		\hline
   		$\sigma(100\omega_\text{b})$ & 0.0038 & - & 0.0036 & 0.0033 & 0.0030 & 0.0029 \\
   		$\sigma( \omega_\text{cdm} )$ & 0.00029 & - & 0.00027 & 0.00026 & 0.00028 & 0.00023 \\
   		$\sigma(100 \theta_s)$ & 0.000085 & - & 0.000085 & 0.000087 & 0.000087 & 0.000086 \\
   		$\sigma (\tau_\text{reio})$ & 0.0021 & - & 0.0020 & 0.0020 & 0.0019 & 0.0018 \\
   		$\sigma (\lnAs)$ & 0.0036 & - & 0.0037 & 0.0035 & 0.0034 & 0.0032 \\
   		$\sigma (n_\text{s})$ & 0.0017 & - & 0.0.0014 & 0.0015 & 0.0014 & 0.0012 \\
   		$\sigma (10^3 n_\text{run})$ & 2.6 & - & 2.3 & 1.5 & 0.41 & 0.30 \\
   		$\sigma (10^9 y)$ & - & - & 0.84 & 0.092 & 0.012 & 0.0016 \\
   		$\sigma (10^8 \mu)$ & - & - & 0.18 & 0.11 & 0.019 & 0.0015 \\
		\hline
	\end{tabular}
	}
\end{center}
\caption{Forecasted $1\sigma$ uncertainties on cosmological parameters for the $\Lambda$CDM model (top) and the $\Lambda$CDM+running of the spectral index (bottom) for different CMB anisotropy experiments (Planck or \lb+\sfour, \lb\ is indicated as "LB" in the table) combined with SD experiments. Here $y$ and $\mu$ are derived quantities inferred from other cosmological parameters, which are tightly constrained by the combination of CMB anisotropies+SDs. As a consequence, $y$ and $\mu$ have much smaller uncertainties than what they would have if evaluated from SDs alone (see text for additional discussions). Note also that the constraints from Planck are calculated via mock likelihoods rather than from the real data. While this makes only a small difference on most of the parameters, it provides a tighter constraint on $\tau_{\rm reio}$ with respect to published results, which, however, does not impact our discussion.}
\vspace{1 cm}
\label{tb_params_err}
\end{table}

While a PIXIE1000 experiment is currently very optimistic, the results presented in this section show that an SD experiment with perfectly controlled systematics and futuristic sensitivity could, in principle, improve the constraints even on just the $\Lambda$CDM model, albeit by a small amount.

Finally, as a remark, note that the SD $y$ and $\mu$ parameters listed in Table \ref{tb_params_err} have been calculated from the cosmological parameters estimated from the MCMC as explained in Section~\ref{sec_class}. Their errors are not to be considered as the error on a direct measurement of the distortion signal, since they are derived from the uncertainties on the cosmological parameters, so that $y$ and $\mu$ can have much smaller uncertainties than what they would have if evaluated considering the corresponding SD mission alone. Therefore, these results translate to model-dependent constraints on $y$ and $\mu$, and only when the SD experiments become sensitive at a level comparable to the given CMB anisotropy mission their combination can effectively improve constraints also on cosmological parameters.

\subsection{Running of the spectral index}\label{sec_running}
In the second scenario that we consider, we investigate the impact that SDs might have on an extension of the $\Lambda$CDM model involving the running of the spectral index, as described in Section~\ref{sec_th_LCDM}. 

\begin{figure}
    \centering
 \includegraphics[width=10 cm]{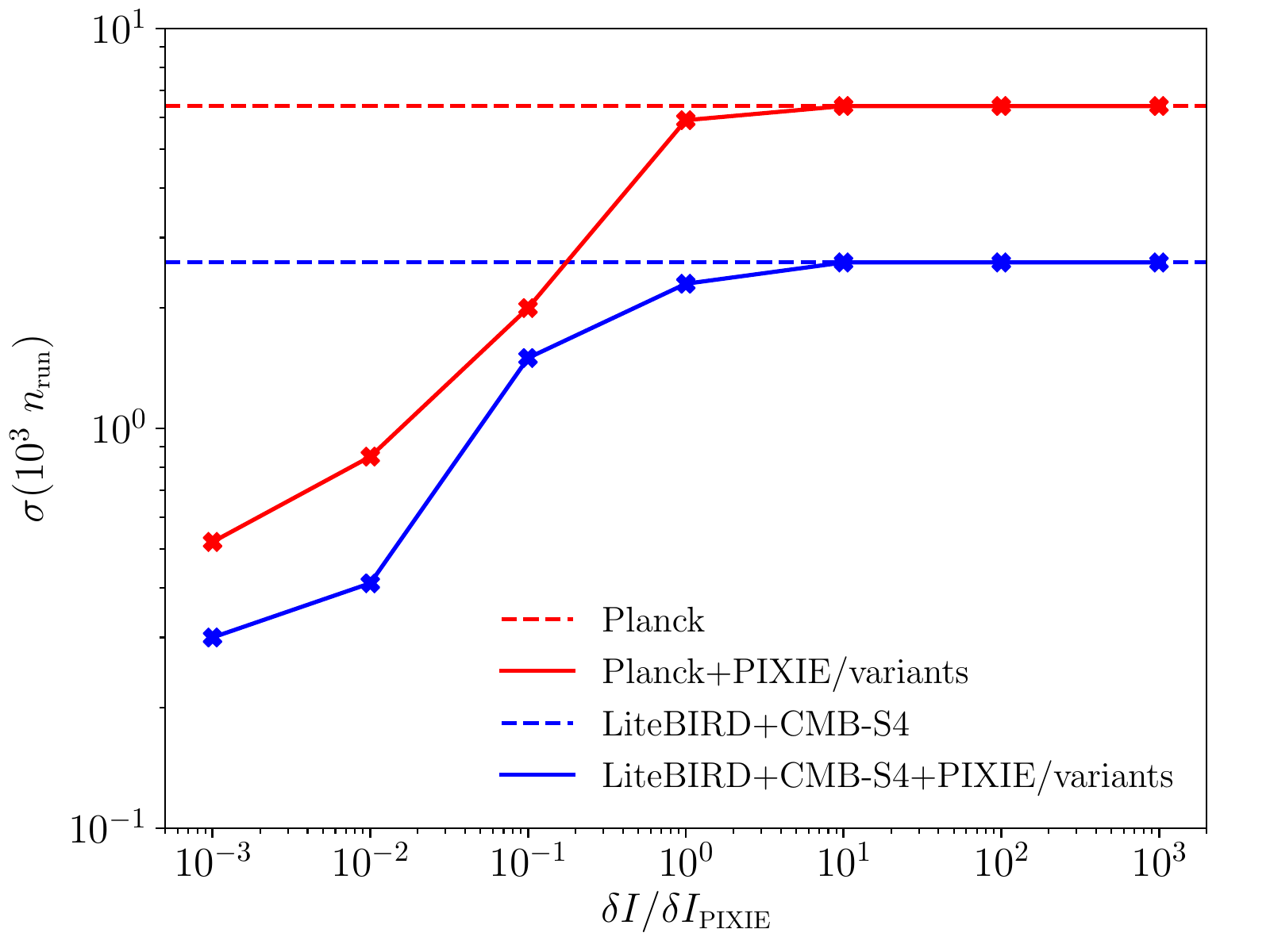}
 \caption{Forecasted $1\sigma$ uncertainties on the running of the scalar spectral index $n_\text{run}$ for different experimental setups, as a function of the PIXIE/variants sensitivity expressed in units of the PIXIE sensitivity, $\delta I/\delta I_{\rm PIXIE}$. The solid red curve shows results for Planck+PIXIE/variants, while the solid blue one shows the case of \lb+\sfour+PIXIE/variants. As a reference, we also show the constraints obtained with Planck (red dashed), as well as the forecasts for \lb+\sfour\ (blue dashed). The constraints of the Planck+FIRAS combination are not explicitly shown as they nearly perfectly overlap with the Planck alone case.}
 \label{fg_sigma_n_run}
\end{figure}

In Figure \ref{fg_sigma_n_run} we show the resulting forecasted uncertainties on the running of the spectral index for the CMB anisotropy experiments Planck or \lb+\sfour\ combined with the SD experiments FIRAS, PIXIE or PIXIE variants. The corresponding numerical values are reported in the lower section of Table \ref{tb_params_err}. For similar and complementary discussions see e.g. Figure 3 of \cite{Chluba:2019nxa} and the related text as well as \cite{Cabass:2016giw}.

The results displayed in Figure \ref{fg_sigma_n_run} allow a number of interesting considerations. First of all, it is clear that the addition of FIRAS to Planck does not improve the bounds on $n_{\rm run}$ at all, while the addition of PIXIE does so only marginally. However, we find that an experiment 10 times more sensitive than PIXIE, such as PRISM, could provide a major improvement on these constraints. When combined with either Planck or \lb+\sfour, PRISM would straighten the bounds by a factor of 3 and 1.7, respectively, with respect to CMB anisotropy data alone, providing a constraint of $\sigma(n_{\rm run}=2\times10^{-3})$ for Planck+PIXIE10  compared to $\sigma(n_{\rm run}=6.5\times10^{-3})$ for Planck alone, or $\sigma(n_{\rm run}=1.5\times10^{-3})$  for \lb+\sfour+PIXIE10 compared to $\sigma(n_{\rm run}=3.5\times10^{-3})$ for \lb+\sfour\footnote{Note that this considerable improvement is due to the fact that the pivot scale has been set very close to the recombination scale, so that the presence of SDs constitute a second reference point very far from that scale, thus greatly improving the measurement. Shifting the pivot scale to higher $k$ values would invert the role of anisotropies and SDs, and the relative improvement on the measurement of $n_{\rm run}$.}. This is potentially very interesting, since such combinations of experiments would start probing values of the running in agreement with the simplest inflationary models, predicting ${|n_{\rm run}|\sim (n_{\rm s}-1)^2\sim 10^{-3}}$~\cite{kosowsky95}. In the presence of foregrounds, which are expected to worsen the sensitivity of SDs to cosmological parameters by a factor of approximately 10 \cite{abitbol_2017}, this expectation would still be within the reach of a PIXIE1000-like mission.

In Table \ref{tb_params_err_2} we also checked the impact of only letting $n_{\rm run}$ vary while fixing all the other cosmological parameters to the best-fit values from \cite{planck_2018_cosmo_params} for the test-case of PIXIE1000. In this case, the constraint on $n_{\rm run}$ would improve by more than one order of magnitude with respect to the combination \lb+\sfour+PIXIE1000 when marginalizing over all parameters. This highlights the importance of performing these forecasts in a consistent way by marginalizing over all of the parameters which impact the SD signal, and by breaking degeneracies by combining with other probes, in our case CMB anisotropies.    
\begin{table}
   	\begin{center}
   		\begin{tabular}{ | c | c | c | c |}
       		\hline\rule{0pt}{2.0ex}
    	     & Planck & PIXIE1000 & Planck+PIXIE1000 \\
       		\hline\rule{0pt}{2.0ex}
       		$\sigma (10^3 n_\text{run})$ & 3.8 & 0.018 & 0.019 \\
       		$\sigma (10^9 y)$ & - & 0.0010 & 0.0011 \\
       		$\sigma (10^8 \mu)$ & - & 0.0015 & 0.0015 \\
       		\hline
   		\end{tabular}
   	\end{center}
   	\caption{Forecasted $1\sigma$ uncertainties on the running of the spectral index by keeping the $\Lambda$CDM parameters fixed to Planck best-fitting values.}
    \label{tb_params_err_2}
\end{table}

\begin{figure}
 \centering
 \includegraphics[width=7 cm, height=7 cm]{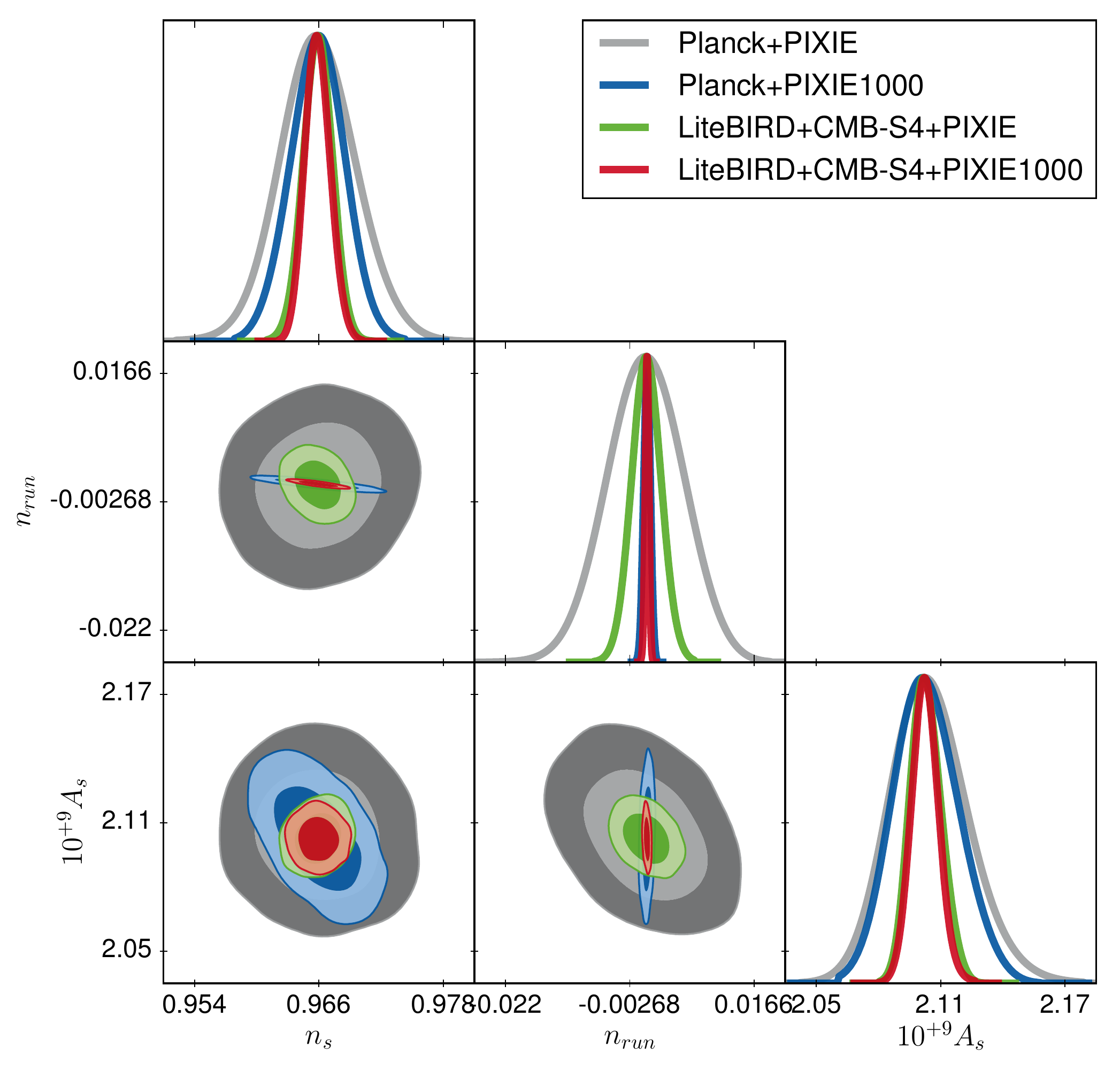}
 \includegraphics[width=5 cm, height=5 cm]{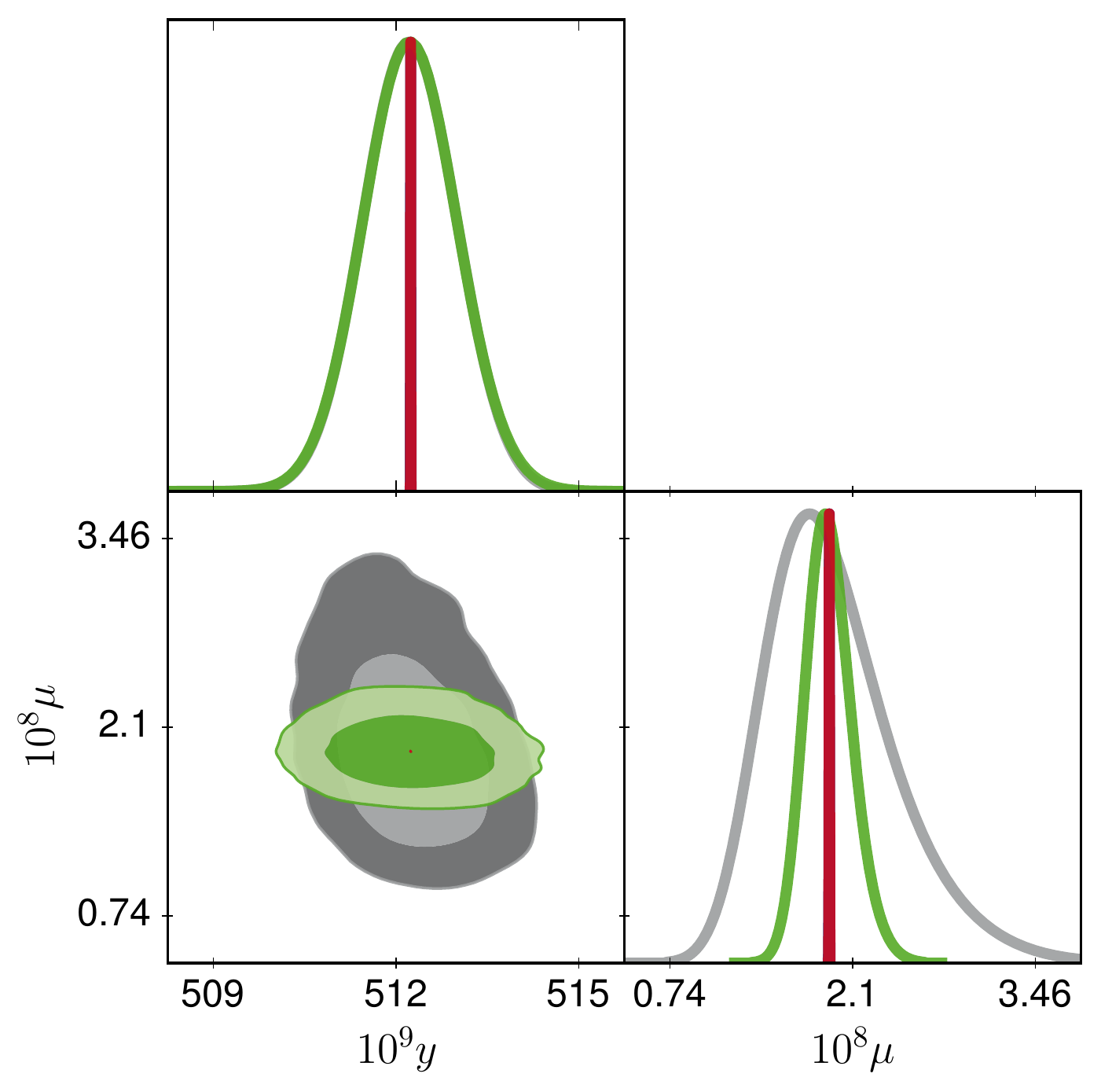}
 \caption{1D posterior distributions and 2D contours (68\% and 95\% C.L.) for the primordial power spectrum amplitude $A_\text{s}$, the spectral index $n_\text{s}$ and the running $n_\text{run}$ (left panel) and the derived $y$ and $\mu$ parameters (right panel) for different experimental configurations.}
 \label{fg_contour_run_y_mu}
\end{figure}

Furthermore, in Figure \ref{fg_contour_run_y_mu} we also show the expected posterior distributions of the power spectrum parameters (left panel), and of the derived $y$ and $\mu$ parameters (right panel) for a selection of the configurations listed in Table \ref{tb_params_err}. It is interesting to notice that when adding an extremely sensitive SD experiment such as PIXIE1000 to a CMB anisotropy experiment such as Planck, a degeneracy appears between the power spectra parameters, in particular $A_s$ and $n_s$, which was absent in the similar Figure 7 of \cite{lucca_2019}. This is due to the fact that the SD experiment becomes sensitive enough to set competitive constraints on the power spectrum amplitude with respect to the anisotropy experiment. However, it cannot completely disentangle a larger amplitude of the overall spectrum $A_s$ from a larger power at very small scales due to an increase in $n_s$, thus imposing a negative correlation between the two parameters.

\subsection{Dark matter annihilation}\label{sec_res_DM_ann}
We now focus on the annihilating DM particle scenario, described by the six $\Lambda$CDM parameters plus the annihilation efficiency $p_\text{ann}$. As in previous cases, we perform forecasts for the cases of Planck and \lb+\sfour\ combined with SD experiments of different sensitivities. Marginalized errors on each parameter are reported in Table \ref{tb_params_err_ann}.
\begin{table}
   	\begin{center}
   	\fontsize{9}{9} \selectfont
	\begin{tabular}{ | c | c | c | c | c | c | c |}
   		\hline\rule{0pt}{2.0ex}
	     & Planck & +FIRAS & +PIXIE & +PIXIE10 & +PIXIE100 & +PIXIE1000 \\
   		\hline\rule{0pt}{2.0ex}
   		$\sigma(100\omega_\text{b})$ & 0.016 & 0.015 & 0.016 & 0.015 & 0.014 & 0.015 \\
   		$\sigma( \omega_\text{cdm} )$ & 0.0013 & 0.0013 & 0.0013 & 0.0012 & 0.00098 & 0.00098 \\
   		$\sigma(100 \theta_s)$ & 0.00034 & 0.00033 & 0.00034 & 0.00034 & 0.00033 & 0.00034 \\
   		$\sigma (\tau_\text{reio})$ & 0.0046 & 0.0045 & 0.0045 & 0.0044 & 0.0039 & 0.0036 \\
   		$\sigma (\lnAs)$ & 0.0093 & 0.0095 & 0.0093 & 0.0088 & 0.0086 & 0.0075 \\
   		$\sigma (n_\text{s})$ & 0.0036 & 0.0036 & 0.0037 & 0.0034 & 0.0015 & 0.0012 \\
   		$\sigma (10^{28} p_{\rm ann})$ & $< 3.6$ & $< 3.5$ & $< 3.4$ & $< 3.1$ & $< 2.4$ & $< 2.1$ \\
   		$\sigma (10^9 y)$ & - & 1476 & 0.83 & 0.083 & 0.010 & 0.0021 \\
   		$\sigma (10^8 \mu)$ & - & 0.085 & 0.083 & 0.068 & 0.014 & 0.0015 \\
   		\hline\rule{0pt}{2.0ex}
	     & LB+CMB-S4 & & & & & \\
   		\hline\rule{0pt}{2.0ex}
   		$\sigma(100\omega_\text{b})$ & 0.0032 & - & 0.0031 & 0.0030 & 0.0028 & 0.0027 \\
   		$\sigma( \omega_\text{cdm} )$ & 0.00027 & - & 0.00028 & 0.00026 & 0.00023 & 0.00021 \\
   		$\sigma(100 \theta_s)$ & 0.000085 & - & 0.000081 & 0.000082 & 0.000081 & 0.000079 \\
   		$\sigma (\tau_\text{reio})$ & 0.0020 & - & 0.0020 & 0.0020 & 0.0017 & 0.0016 \\
   		$\sigma (\lnAs)$ & 0.0035 & - & 0.0035 & 0.0034 & 0.0033 & 0.0030 \\
   		$\sigma (n_\text{s})$ & 0.0014 & - & 0.0014 & 0.0014 & 0.00074 & 0.00042 \\
   		$\sigma (10^{28} p_{\rm ann})$ & $< 1.1$ & - & $< 1.1$ & $< 1.0$ & $< 0.84$ & $< 0.80$ \\
   		$\sigma (10^9 y)$ & - & - & 0.97 & 0.079 & 0.0095 & 0.0013 \\
   		$\sigma (10^8 \mu)$ & - & - & 0.032 & 0.029 & 0.013 & 0.0015 \\
   		\hline
   		\end{tabular}
   	\end{center}
   	\caption{Same as in Table \ref{tb_params_err} but for the annihilating DM case (see Equation \eqref{eq_param_set_ann}). For $p_{\rm ann}$ we show the 95\% C.L. upper limit in units of $\text{cm}^3 / (\text{s GeV})$.
   	}
    \label{tb_params_err_ann}
\end{table}

We find, however, that the addition of SD experiments -- even up to 1000 times the sensitivity of PIXIE -- only improves the bound on $p_{\rm ann}$ by roughly a factor of $1.5$ with respect to Planck or \lb\ plus \sfour\, alone. Therefore, CMB SDs are not a particularly suitable tool to probe the annihilation of relic particles, as already suggested by previous works (see e.g. \cite{chluba_2013}), even when considering the most futuristic SD missions\footnote{Note, however, that our setup relies on the assumption that the annihilation rate stays constant over the whole history of the universe. SDs might still be valuable to test a possible time dependence of the annihilation rate.}.

This is mostly due to two facts. First of all, DM annihilation produces secondary ionizing and Lyman-$\alpha$ photons which can significantly affect the recombination history of the universe, thus leaving a strong imprint on CMB anisotropies. Conversely, such an energy injection would produce a number density of extra photons that would be negligible compared to that of CMB photons, and would result in negligible SDs. Secondly, as can be seen e.g. in Figure 4 of \cite{lucca_2019}, the energy injection due to DM annihilation is nearly constant at redshifts $z>10^4$. Since the same is also true for the dissipation of acoustic waves (see again Figure 4 of \cite{lucca_2019}), the respective effects of $p_{\rm ann}$ and $A_{\rm s}$ on SDs are partly degenerate. Thus $p_{\rm ann}$ alone cannot be efficiently constrained by SD experiments.

\subsection{Dark matter decay}\label{sec_res_DM_dec}
Unlike in the case of DM annihilation, SD observations are an extremely powerful constraining tool in the context of decaying DM. This is because they allow to probe a range of DM lifetimes lying outside of the reach of other complementary probes such as CMB anisotropies.

\begin{figure}
    \centering
 \includegraphics[width=10 cm]{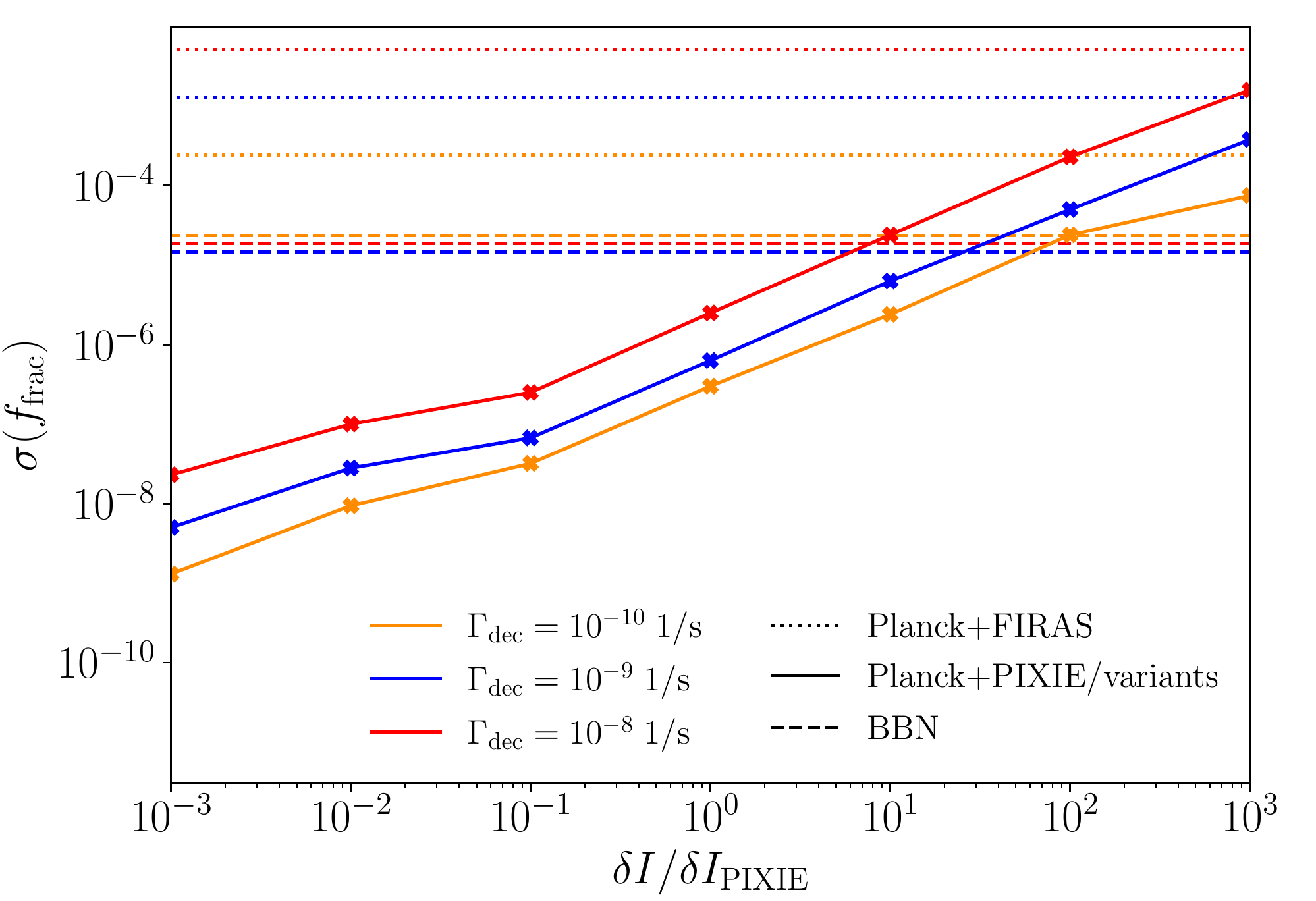}
 \caption{Forecasted 95\% C.L. upper limits on the decaying DM fraction $f_{\rm frac}$ for three different particle lifetimes, for Planck combined with PIXIE variants (solid lines). As a reference, the current bounds from FIRAS are shown as horizontal dotted lines and the BBN bounds from \cite{Poulin:2016anj} as horizontal dashed lines. Note that, differently from what is done in Figure \ref{fg_sigma_n_run}, here we do not show the \lb+\sfour+PIXIE/variants bounds, as they do not differ from the Planck+PIXE/variants already shown.}
 \label{fg_sigma_contour_f_DM}
\end{figure}
In order to precisely evaluate the extent of the constraining power of SDs, we perform forecasts on the parameters influencing the decay of DM particles, i.e. the decaying DM fraction $f_{\rm frac}$ and its lifetime expressed in terms of the decay width $\Gamma_{\rm dec} \equiv 1/\tau_{\rm dec}$, by considering a $6+2$ extension of the $\Lambda$CDM model including $\{f_{\rm frac}, \Gamma_{\rm dec} \}$. To efficiently perform the MCMC runs, we slice the parameter space along $\Gamma_{\rm dec}$ for different discrete values and sample the remaining $6+1$ parameter space -- otherwise the non-convex topology of the posterior distribution in the $f_{\rm frac}-\Gamma_{\rm dec}$ plane would slow down the convergence of the chains and highly increase the CPU time \cite{lucca_2019}.

In Figure \ref{fg_sigma_contour_f_DM} we show the resulting uncertainties on $f_{\rm frac}$ for three different particle decay rates ($\Gamma_{\rm dec}= 10^{-8}$, $10^{-9}$ and ${10^{-10}\text{ 1/s}}$) as a function of the sensitivity of the experiment (solid lines). The same quantities are also listed in Table \ref{tb_params_err_decay}. For reference, in the figure we also display the corresponding FIRAS predictions as dotted lines and the current BBN bounds for the different $\Gamma_{\rm dec}$ as given in \cite{Poulin:2016anj}.
\begin{table}
   	\begin{center}
   	\fontsize{9}{9} \selectfont
	\begin{tabular}{ | c | c | c | c | c | c | c |}
   		\hline 
    	\multicolumn{7}{|c|}{$\Gamma_{\rm dec} = 10^{-8}$ 1/s} \\
   		\hline\rule{0pt}{2.0ex}
	     & Planck & +FIRAS & +PIXIE & +PIXIE10 & +PIXIE100 & +PIXIE1000 \\
   		\hline\rule{0pt}{2.0ex}
   		$\sigma(100\omega_\text{b})$ & 0.015 & 0.015 & 0.015 & 0.015 & 0.015 & 0.015 \\
   		$\sigma( \omega_\text{cdm} )$ & 0.0013 & 0.0013 & 0.0012 & 0.0013 & 0.0011 & 0.00094 \\
   		$\sigma(100 \theta_s)$ & 0.00035 & 0.00035 & 0.00035 & 0.00035 & 0.00034 & 0.00033 \\
   		$\sigma (\tau_\text{reio})$ & 0.0046 & 0.0046 & 0.0046 & 0.0045 & 0.0040 & 0.0034 \\
   		$\sigma (\lnAs)$ & 0.0086 & 0.0089 & 0.0088 & 0.0085 & 0.0084 & 0.0074 \\
   		$\sigma (n_\text{s})$ & 0.0039 & 0.0038 & 0.0036 & 0.0037 & 0.0023 & 0.0011 \\
   		$\sigma (f_\text{frac})$ & - & $<5.1\times10^{-3}$ & $<2.5\times10^{-6}$ & $<2.5\times10^{-7}$ & $<1.0\times10^{-7}$ & $<2.3\times10^{-8}$ \\
   		$\sigma (10^9 y)$ & - & 1039 & 0.92 & 0.094 & 0.011 & 0.0019 \\
   		$\sigma (10^8 \mu)$ & - & 1411 & 0.79 & 0.10 & 0.015 & 0.0016 \\
   		\hline
   		\hline 
    	\multicolumn{7}{|c|}{$\Gamma_{\rm dec} = 10^{-9}$ 1/s} \\
   		\hline\rule{0pt}{2.0ex}
	     & Planck & +FIRAS & +PIXIE & +PIXIE10 & +PIXIE100 & +PIXIE1000 \\
   		\hline\rule{0pt}{2.0ex}
   		$\sigma(100\omega_\text{b})$ & 0.015 & 0.015 & 0.016 & 0.015 & 0.015 & 0.015 \\
   		$\sigma( \omega_\text{cdm} )$ & 0.0013 & 0.0013 & 0.0013 & 0.0012 & 0.0011 & 0.00099 \\
   		$\sigma(100 \theta_s)$ & 0.00035 & 0.00034 & 0.00033 & 0.00034 & 0.00033 & 0.00033 \\
   		$\sigma (\tau_\text{reio})$ & 0.0046 & 0.0044 & 0.0044 & 0.0044 & 0.0039 & 0.0033 \\
   		$\sigma (\lnAs)$ & 0.0086 & 0.0085 & 0.0087 & 0.0086 & 0.0080 & 0.0070 \\
   		$\sigma (n_\text{s})$ & 0.0039 & 0.0037 & 0.0038 & 0.0037 & 0.0023 & 0.0017 \\
   		$\sigma (f_\text{frac})$ & - & $<1.3\times10^{-3}$ & $< 6.3 \times 10^{-7}$ & $< 6.7 \times 10^{-8}$ & $<2.8\times10^{-8}$ & $<5.0\times10^{-9}$ \\
   		$\sigma (10^9 y)$ & - & 1186 & 0.88 & 0.089 & 0.011 & 0.0025 \\
   		$\sigma (10^8 \mu)$ & - & 1394 & 0.75 & 0.093 & 0.015 & 0.0015 \\
        \hline
   		\hline 
    	\multicolumn{7}{|c|}{$\Gamma_{\rm dec} = 10^{-10}$ 1/s} \\
   		\hline\rule{0pt}{2.0ex}
	     & Planck & +FIRAS & +PIXIE & +PIXIE10 & +PIXIE100 & +PIXIE1000 \\
   		\hline\rule{0pt}{2.0ex}
   		$\sigma(100\omega_\text{b})$ & 0.015 & 0.015 & 0.015 & 0.014 & 0.015 & 0.014 \\
   		$\sigma( \omega_\text{cdm} )$ & 0.0013 & 0.0012 & 0.0013 & 0.0011 & 0.0010 & 0.00091 \\
   		$\sigma(100 \theta_s)$ & 0.00033 & 0.00035 & 0.00034 & 0.00034 & 0.00035 & 0.00031 \\
   		$\sigma (\tau_\text{reio})$ & 0.0046 & 0.0046 & 0.0045 & 0.0040 & 0.0039 & 0.0036 \\
   		$\sigma (\lnAs)$ & 0.0086 & 0.0085 & 0.0085 & 0.0083 & 0.0085 & 0.0077 \\
   		$\sigma (n_\text{s})$ & 0.0039 & 0.0037 & 0.0038 & 0.0033 & 0.0018 & 0.00092 \\
   		$\sigma (f_\text{frac})$ & - & $<2.4\times10^{-4}$ & $<3.0\times10^{-7}$ & $<3.2\times10^{-8}$ & $<9.4\times10^{-9}$ & $<1.3\times10^{-9}$ \\
   		$\sigma (10^9 y)$ & - & 1478 & 0.99 & 0.089 & 0.010 & 0.0019 \\
   		$\sigma (10^8 \mu)$ & - & 662 & 0.77 & 0.096 & 0.015 & 0.0014 \\
   		\hline
   		\end{tabular}
   	\end{center}
   	\caption{Same as in Table \ref{tb_params_err} but for the decaying DM case (see Equation \eqref{eq_param_set_dec}) for different DM lifetimes. For $f_{\rm frac}$ we show the 95\% C.L. upper limit. For the Planck alone case, which is not sensitive to the lifetimes considered here, we just report for reference the $\Lambda$CDM constraints already listed in Table \ref{tb_params_err}.}
    \label{tb_params_err_decay}
\end{table}

The results shown in Figure \ref{fg_sigma_contour_f_DM} rely on the combination of future SD experiments with Planck data. CMB anisotropies are know to be directly sensitive to DM decay only when the lifetime is larger than ${\cal O}(10^{12}\text{~s})$, such that the released energy may affect the thermal history of the universe around the time of recombination. Still, for the smaller lifetimes considered here, the inclusion of future CMB anisotropy data could in principle strengthen the bounds on $\{f_{\rm frac}, \Gamma_{\rm dec} \}$ through the reduction of parameter degeneracies. We performed dedicated runs with Planck replaced by \lb+\sfour{} and found that this is not the case. The reason is that the parameters $\{f_{\rm frac}, \Gamma_{\rm dec} \}$, that are probed only by SD experiments for $\Gamma_{\rm dec}\geq10^{-10}$ 1/s, do not appear to be degenerate with any other $\Lambda$CDM parameter.

A key conclusion of this section is that a future SD mission with a sensitivity one order of magnitude worse than PIXIE would already be able to set stronger bounds on the decaying DM fraction than BBN (true even accounting for the drop in sensitivity expected due to the presence of foregrounds). This can be quantitatively seen comparing the solid (SDs) and the dashed (BBN) lines in Figure \ref{fg_sigma_contour_f_DM}. This means that, even using current technology, SD experiments could be by far the best available probe of the thermal history prior to recombination. This is even more true considering that, as already argued in \cite{lucca_2019}, we are reaching the maximum amount of information that can be extracted from BBN observables, and therefore significant improvements in the corresponding sensitivities are not expected. 

As a final remark, note that in Table \ref{tb_params_err_decay} the constraints on several $\Lambda$CDM parameters also improve once more accurate SD data are included. This shows that even in the case of DM decay, there is enough information in the SD signal to still improve the constraints on multiple parameters such as $\lnAs$ and $n_s$.

\subsection{Feasibility in different environments\label{sec:feasibility}}
In the previous sections we explored the synergies that future SD experiments could potentially have with CMB anisotropy experiments. 
However, most of the experimental configurations considered in the previous forecasts assume observations from a satellite in a very wide range of frequencies (see Section \ref{sec_pixie}). Realistically, such experiments will be preceded by pathfinders from the ground or from a balloon which will allow us to test the required technologies. For such forerunners, the atmosphere represents a formidable contaminant \cite{kuo}. First, photon absorption in water vapor limits the observable frequency windows. Second, the average brightness of the sky increases the level of photon noise in the detectors. Third, the dependence of this brightness over time and direction introduces further systematics and compromises the efficiency of noise subtraction.

In this section, we explore the ability of a ground-based or balloon-borne experiment to measure an SD signal. We consider the experimental configurations described in Section~\ref{sec_exp_diff_env}. As mentioned there, we take into account two of the main limitations induced by the atmosphere, i.e. narrower frequency bands and an increased photon noise due to the average brightness of the sky. However, we do not take into account the third and potentially most problematic source of contamination, the fluctuations in sky brightness. Strategies to mitigate the corresponding uncertainties are in development (see e.g. \cite{Errard:2015twg}), and we leave the analysis of their impact for future work.

\begin{figure}
\centering
\includegraphics[width=12 cm]{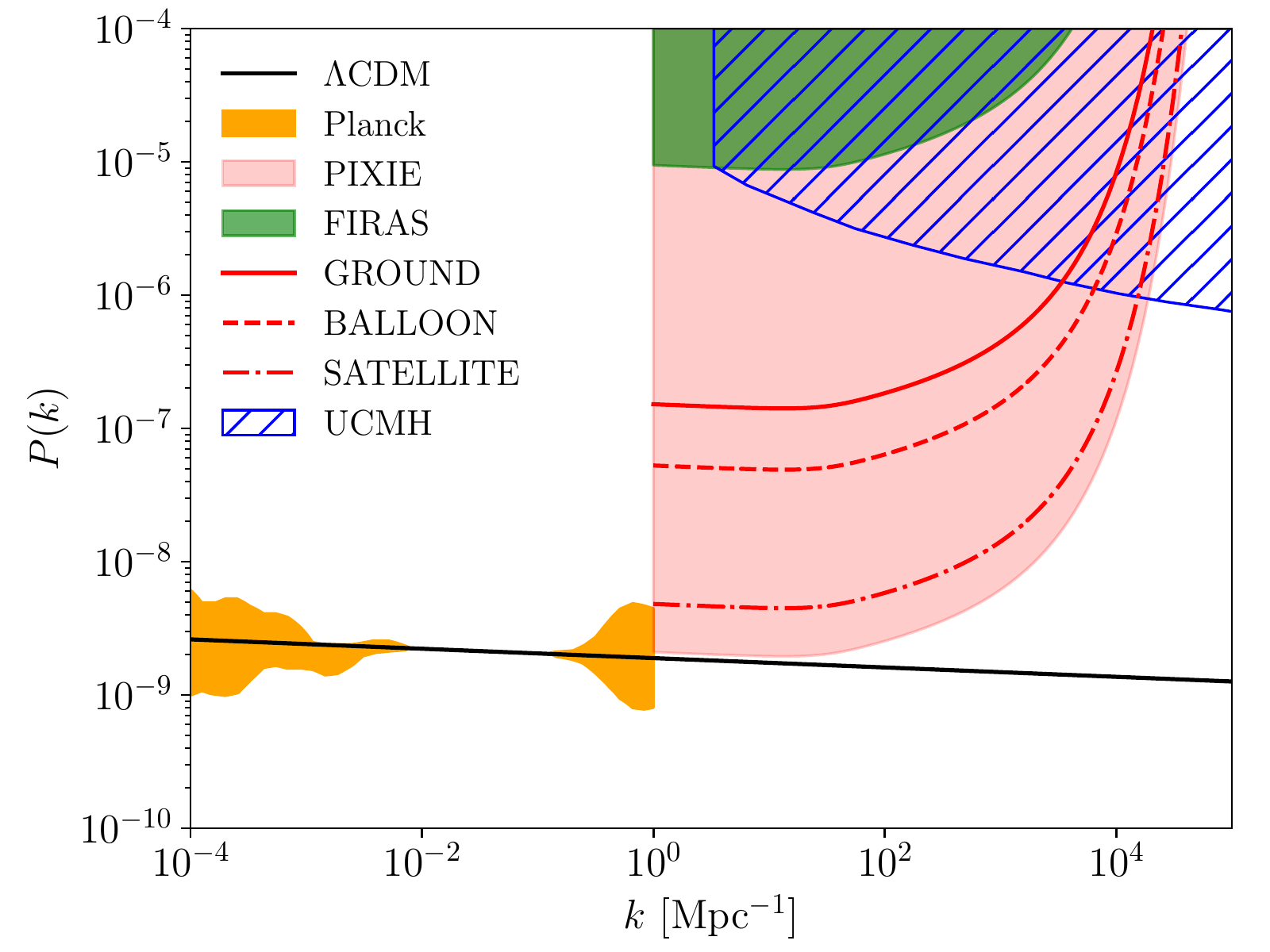}
\caption{Current and forecasted constraints on the primordial power spectrum shape. As a reference, the black solid line shows the power-law $\Lambda$CDM prediction (assuming Planck best-fitting values of $A_s$ and $n_s$). The orange region corresponds to the 2$\sigma$ uncertainty on the primordial power spectrum inferred from Planck CMB anisotropy data (similarly to Figure 24 of \cite{Akrami:2018odb}). For simplicity, the Planck constraint has been cut at 1 Mpc$^{-1}$. The dashed blue area represents the region of parameter space that can be excluded by the non-observation of Ultra-Compact Mini Halos (UCMH) \cite{Bringmann:2011ut}. The green and pink regions correspond to the area of parameter space probed respectively by FIRAS (${y<1.5 \times 10^{-5}}$ and $\mu<9 \times 10^{-5}$ at 2$\sigma$, \cite{fixsen_1996}) and PIXIE ($y\simeq 4 \times 10^{-9}$ and $\mu\simeq 2 \times 10^{-8}$ at 2$\sigma$, \cite{kogut_2011}). The curves have been calculated according to the analytical approximations presented in \cite{chluba_grin_2013} (see in particular Equations (24)-(25) therein). As the approximations are only valid for $k>1$ Mpc$^{-1}$, the contours are sharply cut a that scale. For a graphical comparison, similar contours have already been shown e.g. in Figure 4 of \cite{Chluba:2012we}. Finally, the red curves represent the regions potentially probed at 2$\sigma$ by the ground-based (solid), balloon (dashed) and satellite (dot-dashed) configurations considered in this work.}
\label{fg_Pk}
\end{figure}

Here we study the case of the $\Lambda$CDM model, in which SDs are produced by the dissipation of acoustic waves and baryon cooling. For this model, the ground-based or balloon-borne experiments considered here are not sensitive enough to set competitive bounds on $\{A_s, n_s\}$ compared to CMB anisotropy data. However, we should stress again that such new SD bounds, although not very strong, would provide independent information on the primordial power spectrum at scales $k\sim 1-10^4$ Mpc$^{-1}$, and thus give some precious insight on possible deviations from a power-law or running primordial spectrum on these scales (see Figure \ref{fg_Pk} for a graphical overview).

Therefore, in this Section we will only use SD experiments to set a constraint on $A_s$, while assigning Gaussian priors derived from the Planck experiment to all other parameters. Furthermore, for these runs we will also marginalize over the scattering optical depth $\Delta\tau$ and the electron temperature $T_e$, as discussed in Section \ref{sec_montepython}, in order to account for uncertainties on the SD signal caused by the ICM, which provides the largest late-time contribution to the total SD signal \cite{hill_2015}. Indeed, the level of this signal will not be accurately known prior to such experiments. On the contrary, this will constitute one of their main targets. However, as in the previous sections, here we also neglect the deterioration of the signal due to the presence of galactic and extra-galactic foregrounds. Although a precise estimate of the impact that such foregrounds would have on the experimental configurations considered in this section is not present in the literature, we expect them to degrade the final sensitivity to the SD signal by a factor larger than 10 (which was derived for PIXIE \cite{abitbol_2017}) because of the limited frequency array. We leave a more detailed analysis of this aspect for future work.

Our results for a ground-based, balloon and satellite configuration are presented in Table~\ref{tb_diff_env_res}. In order to quantify the impact of marginalizing over late-time sources we also quote the results obtained without marginalizing over late-time sources for the satellite configuration. As clearly shown in the table, a detection of the ICM signal ($y\sim 10^{-6}$) would be in the reach of a balloon experiment (with $\sigma(y)\simeq 2\times10^{-8}$), while the primordial $\Lambda$CDM signal, although with a low significance, would only be possible with an SD mission from space (with ${\sigma(A_s)\simeq2\times10^{-9}}$). A ground-based experiment would instead only be able to set an upper bound on $A_s$ and $y$, as has been the case for FIRAS.
\begin{table}
\begin{center}
	\begin{tabular}{| c | c | c | c | c |}
		\hline
		 & GROUND & BALLOON & SATELLITE & SATELLITEnoreio \\
		\hline 
		$\sigma (10^9 A_\text{s})$ & $< 137$ & $< 47.1$ & 2.2 & 0.23 \\
		$\sigma (10^9y)$ & $<2616$ & 20.6 & 15.8 & 0.43 \\
		$\sigma (10^8\mu)$ & $<145$ & $<50.3$ & 2.3 & 0.25 \\
		\hline
	\end{tabular}
\end{center}
\caption{Forecasted 2$\sigma$ upper bounds and $1\sigma$ uncertainties on the power spectrum amplitude, as well as on the $y$ and $\mu$ parameters, for the experimental setups in different environments. Note that we have set Gaussian priors derived from the Planck experiment on all the cosmological parameters except for $A_{\rm s}$. The fiducial value of the primordial amplitude parameter is ${A_s = 2.106 \times 10^9}$. The configurations labeled SATELLITEnoreio refer to the same setup as SATELLITE, but without marginalizing over the contribution of reionization and late-time sources.}
\label{tb_diff_env_res}
\end{table}

However, in order to truly appreciate the constraining power of the experimental configurations discussed in this Section, it is useful to compare them with current constraints from other types of observations. For this purpose, we show in Figure \ref{fg_Pk} the most up-to-date constraints on the primordial power spectrum shape, coming from a variety of complementary probes at different scales. In particular, we compare the bounds imposed by Planck at large scales with those from several SD observations at small scales. While the former are inferred from Planck CMB anisotropy data (similarly to Figure 24 of \cite{Akrami:2018odb}), the latter have been calculated according to the analytical approximations presented in \cite{chluba_grin_2013} (see in particular Equations (24)-(25) therein). The red lines represent the different environments considered in this section, while for the sensitivities of FIRAS and PIXIE we refer to the results obtained in \cite{fixsen_1996} and \cite{kogut_2011}, respectively\footnote{We do not consider the values presented in Sec. \ref{sec_lcdm_params} because they are mainly driven by the inclusion of Planck for FIRAS and PIXIE, and we refrain from showing more realistic forecasts such as the ones presented in \cite{abitbol_2017} which also include foregrounds (while the ones we employ based on \cite{kogut_2011} do not) for sake of consistency with the other configurations we display.}.

The shape of the SD bounds follows the extrapolated $\Lambda$CDM prediction up to $k$ values of the order of 25 Mpc$^{-1}$. In this range, the bounds match the uncertainties on $A_s$ reported in Table \ref{tb_diff_env_res}. For higher wavenumbers the constraining power of SDs rapidly deteriorates because of the exponential drop in the production of $\mu$ distortions at redshifts of the order of $10^6$ (proportionally to the visibility function, see e.g. \cite{danese_de_zotti, hu_1994}, as well as Figure 2 of~\cite{chluba_jeong_2014}).

Focusing now on the red lines representing the configurations listed in Table \ref{tb_diff_env_res}, it becomes clear from Figure \ref{fg_Pk} that even the least sensitive of the considered setups would improve by up to two orders of magnitude the current most stringent bounds on the primordial power spectrum at scales higher than $k\simeq 1$ Mpc$^{-1}$. The improvement extends over four decades in $k$ space and, although with a reduced significance, can be expected to still be present even after the inclusion of galactic and extra-galactic foregrounds. This clearly shows that a future SD mission, even from the ground, would greatly contribute to our understanding of the inflationary epoch at scales yet unexplored by any other cosmological probe.

Finally, Table \ref{tb_diff_env_res} also shows the impact of not marginalizing over the late-time effects for the satellite case. In this case, it becomes possible to use the information stored in the $y$ distortion to constrain the early universe. In the $\Lambda$CDM case, this would give an opportunity to tighten the constraint on $A_{\rm s}$ by a factor of $10$.

\section{Conclusion}\label{sec_concl}
The recent advent of precision cosmology has enabled us to test the $\Lambda$CDM model and its extensions with an unprecedented level of accuracy. In particular, CMB anisotropies have provided very stringent constraints on a wide range of cosmological models. In order to further explore and test these scenarios, complementary measurements are required. In this paper we studied the possibility of combining CMB anisotropy experiments with CMB SD measurements, extending the work of \cite{lucca_2019}.

SDs are predicted to exist even within the standard $\Lambda$CDM scenario and can in principle help to constrain its free parameters. For instance, SDs provide unique information on the shape of the primordial scalar power spectrum at scales much smaller than those probed by CMB anisotropies. Furthermore, SDs are also sensitive to more exotic models which modify the thermal history of the universe. Therefore, they can constrain energy injection phenomena which happened at times much earlier than the epoch of recombination, such as the early decay of DM particles.

Within this work, we investigated the amount of information that will be possible to extract from the combination of CMB anisotropies and SD measurements, even in very futuristic SD configurations. To this purpose, based on the setups of proposed CMB SD experiments such as PIXIE, we explored the constraining power of combining current and up-coming CMB anisotropy missions -- such as Planck or \lb\ plus \sfour\ -- with SD experiments with different sensitivities. This allowed us to present constraints on some of the most interesting parameters which impact SDs (such as the running of the scalar spectral index), while consistently marginalizing over the uncertainties on other cosmological parameters. Indeed, parameters such as the baryon or the DM density would be almost completely degenerate in an analysis of SD experimental data alone, but can be strongly constrained by CMB anisotropy experiments. These forecasts marginalize over late-time sources of SDs such as the ICM. However, they assume the perfect removal of other galactic and extra-galactic foregrounds.

For the $\Lambda$CDM case, we find that an experiment 1000 times more sensitive than PIXIE, comparable to the proposed Voyager 2050 mission, would in principle be able to improve the constraints on $A_{\rm s}$ and $n_{\rm s}$ inferred from \lb+\sfour\ data alone, albeit by a small amount. On the other hand, we find that for the running of the spectral index an experiment 10 (resp. 1000) times more sensitive than PIXIE combined with \lb+\sfour\ could improve the constraints on $n_{\rm run}$ by a factor of almost 2 (9) with respect to the anisotropy data alone. Moreover, DM annihilation constraints would only improve by up to a factor of 1.4 with respect to anisotropy experiments alone even with an SD experiment 1000 times more sensitive than PIXIE. On the contrary, when exploring models of DM decay with lifetimes shorter than the age of universe at the time of recombination, SD experiments with sensitivities already 10 times worse than PIXIE could provide constraints stronger than current CMB and BBN measurements. 

Finally, we discuss a few short-term SD experiments, which could be considered as pathfinders for more futuristic configurations. In particular, we consider a ground-based and a balloon configuration, which we then compare with a similar setup sent to space, i.e. a satellite. Also in this case, even if the sensitivities are not competitive with current CMB anisotropy missions, SDs can provide important insights to our understanding of the inflationary epoch. In fact, although a direct detection of the primordial $\Lambda$CDM SD signal would be out of reach, even a measurement from the ground could potentially set the currently strongest constraints on the amplitude of the primordial power spectrum at scales between 1~Mpc$^{-1}$ and 10$^{4}$ Mpc$^{-1}$. In addition to that, the detection of the late-time SD signal (possible already with a balloon) would also considerably help in understanding the epoch of reionization and structure formation. Again, these are optimistic forecasts, since foreground cleaning as well as atmospheric brightness fluctuations will represent a formidable source of contamination for these measurements, calling for more accurate and realistic forecasts in the future. 

Overall, with respect to \cite{lucca_2019}, we explored a much larger range of experimental configurations, from ground-based to very futuristic satellite experiments, exploring their complementarity with CMB anisotropies and explicitly accounting for the impact of marginalizing over the contribution of the ICM. Furthermore, we calculated  constraints for the $\Lambda$CDM model, which in futuristic scenarios could potentially be interesting.

In conclusion, SDs are a potentially competitive cosmological tool. Although a direct detection of the $\Lambda$CDM SD signal is still missing, we clearly showed in this work that their constraining power ranges from greatly improving existing constraints to probing yet unexplored cosmological scales relevant, for instance, for our understanding of inflation and DM.

\section*{Acknowledgements}

We thank Jens Chluba for reading the manuscript and providing very helpful suggestions.
HF acknowledges support from the IAP for the completion of his Master's work and supplying computational resources.
ML is supported by the ''Probing dark matter with neutrinos'' ULB-ARC convention and the F.R.S./FNRS under the Excellence of Science (EoS) project No. 30820817 - be.h ''The H boson gateway to physics beyond the Standard Model”. DH is supported by the FNRS research grant number \mbox{F.4520.19}. JL acknowledges support for this project from the DFG grant \mbox{LE~3742/3-1}. NS acknowledges support from the DFG grant \mbox{LE~3742/4-1}.

\bibliography{my_bibliography}
\bibliographystyle{JHEP}

\end{document}